\documentclass[pre]{revtex4}

\usepackage{graphicx}
\usepackage{dcolumn}
\usepackage{bm}
\usepackage{subfigure}
\usepackage{amssymb}
\usepackage{graphics}
\usepackage{epsfig}
\usepackage{threeparttable}

\usepackage{threeparttable}

\begin{document}

\preprint{APS/123-QED}

\title{Investigation of bond dilution effects on the magnetic properties of a cylindrical Ising nanowire }
\author{Yusuf Y\"{u}ksel}
\altaffiliation[Also at ]{Dokuz Eyl\"{u}l University,
Graduate School of Natural and Applied Sciences, Turkey}
\author{\"{U}mit Ak{\i}nc{\i}}%
\affiliation{Department of Physics, Dokuz Eyl\"{u}l University,
TR-35160 Izmir, Turkey}
\author{Hamza Polat}
\email{hamza.polat@deu.edu.tr}
\affiliation{Department of Physics, Dokuz Eyl\"{u}l University,
TR-35160 Izmir, Turkey}


\date{\today}

\begin{abstract}
A cylindrical magnetic nanowire system composed of ferromagnetic core and shell layers has been investigated by using effective field theory with correlations. Both ferromagnetic and antiferromagnetic exchange couplings at the core-shell interface have been considered. Main attention has been focused on the  effects of the quenched disordered shell bonds, as well as interface bonds on the magnetic properties of the system. A complete picture of the phase diagrams and magnetization profiles has been represented. It has been shown that for the antiferromagnetic nanowire system, the magnetization curves can be classified according to N\'{e}el theory of ferrimagnetism and it has been found that under certain conditions, the magnetization profiles may exhibit Q-type, P-type, N-type and L-type behaviors. The observed L-type behavior has not been reported in the literature before for the equilibrium properties of nanoscaled magnets. As another interesting feature of the system, it has been found that a compensation point can be induced by a bond dilution process in the surface. Furthermore, we have not found any evidence of neither the first order phase transition characteristics, nor the reentrance phenomena.
\end{abstract}

\maketitle

\section{Introduction}\label{introduction}
Nowadays, magnetic nanostructures i.e. nanoparticles such as nanorods, nanobelts, nanowires and nanotubes have drawn considerable attention from both theoretical and experimental points of view, due to their technological \cite{kim,kodama} and biomedical \cite{pankhurst,habib,sounderya} applications such as information storage devices and drug delivery in cancer thermotherapy. In general, these fine particles are used for manufacturing biosensors \cite{kurlyandskaya} and permanent magnets \cite{zeng}. Recent developments in the experimental techniques allow the scientists to fabricate such kinds of fine nanoscaled materials \cite{ruhrig,schrefl}, and the magnetization of certain nanomaterials such as $\mathrm{\gamma-Fe_{2}O_{3}}$ nanoparticles has been experimentally measured \cite{martinez}. In particular, magnetic nanowires and nanotubes such as $\mathrm{ZnO}$ \cite{fan}, $\mathrm{FePt}$ and $\mathrm{Fe_{3}O_{4}}$ \cite{su} can be synthesized by various experimental techniques and they have many applications in nanotechnology \cite{skomski,schlorb}, and they are also utilized as raw materials in fabrication of ultra-high density magnetic recording media \cite{wegrowe,fert,bader}.

On the theoretical side, much effort has also been devoted, and these systems have been studied by a wide variety of techniques such as mean field theory (MFT) \cite{leite,kaneyoshi0,kaneyoshi1,kaneyoshi2,kaneyoshi3,kaneyoshi5,kaneyoshi9}, effective field theory (EFT) with correlations \cite{kaneyoshi2,kaneyoshi3,kaneyoshi5,kaneyoshi9,kaneyoshi4,kaneyoshi6,kaneyoshi7,canko1,wang,keskin,kaneyoshi8,sarli,canko2,jiang1,kaneyoshi10,bouhou}, Green functions (GF) formalism \cite{garanin}, variational cumulant expansion (VCE) \cite{wang1,wang2}, and Monte Carlo (MC) simulations \cite{eftaxias,iglesias3,kechrakos,hu1,trohidou1,wu,iglesias1,iglesias2,trohidou2,vasilakaki,hu2,du,zaim1,zaim2,jiang2,yuksel}. Monte Carlo simulation \cite{metropolis} is regarded as a powerful numerical approach for simulating the behavior of many complex systems, including magnetic nanoparticle systems. For instance, by utilizing this tool, Refs. \cite{eftaxias,iglesias3,kechrakos,hu1,trohidou1,wu,iglesias1,iglesias2,trohidou2,vasilakaki,hu2,du} investigated the exchange bias effect in magnetic core-shell nanoparticles where the hysteresis loop exhibits a shift below the Ne\'{e}l temperature of the antiferromagnetic shell due to the exchange coupling at the interface region of ferromagnetic core and antiferromagnetic shell. Furthermore, according to recent Monte Carlo studies, it has been shown that the core-shell concept can be successfully applied in nanomagnetism since it is capable of explaining various characteristic behaviors observed in nanoparticle magnetism \cite{zaim1,zaim2,jiang2,yuksel}. On the other hand, EFT which is superior to conventional MFT has been introduced for the first time by Kaneyoshi for ferroelectric nanoparticles \cite{kaneyoshi2}. In a series of the consecutive studies, he extended the theory for the investigation of thermal and magnetic properties of the nanoscaled transverse Ising thin films \cite{kaneyoshi9}, and also for cylindrical nanowire and nanotube systems \cite{kaneyoshi3,kaneyoshi5,kaneyoshi4,kaneyoshi6,kaneyoshi7,kaneyoshi8,kaneyoshi10}.

Magnetic nanoparticles serve as a bridge between bulk materials and atomic or molecular structures. Physical properties of a bulk material are independent from size of the material; however, below a critical size, nanoparticles often exhibit size-dependent properties and some unique phenomena such as superparamagnetism \cite{kittel,jacobs}, quantum tunneling of the magnetization \cite{chudnovsky}, and unusual large coercivities \cite{kneller} have been reported. As an example, it has been experimentally shown that $\mathrm{La_{0.67}Ca_{0.33}MnO_{3}}$ (LCMN) nanoparticle exhibits a negative core-shell coupling, although the bulk LCMN is a ferromagnet \cite{bhowmik1,bhowmik2}. Moreover, as a theoretical example, the total magnetizations in a nanoscaled transverse Ising thin film with thickness $L$ are investigated by the use of both the EFT with correlations and the MFT, and it has been shown that the magnetization may exhibit two compensation points with the increasing film thickness \cite{kaneyoshi9}. The phenomenon of the observation of two compensation points in the nanoscaled thin films has also been reported for bulk ferrimagnetic materials \cite{bobak,xin,cekiz1,kaneyoshi12,cekiz2}. However, the origin of the existence of such a compensation point in the nanoscaled magnets are quite different from those observed in the bulk ferrimagnetic materials. Namely, a compensation point originates in the bulk systems due to the different temperature dependence of the atomic moments of the sublattices. However, nanoscaled magnetic particles such as nanowires or nanotubes exhibit a compensation point, due to the presence of an antiferromagnetic interface coupling between the core and the shell, even if the lattice sites in the particle core and shell are occupied by identical atomic moments. Hence, theoretical investigation of ferrimagnetism in nanoparticle systems has opened a new field in the research of the critical phenomena in nanoscaled magnetic particles \cite{kaneyoshi1}.

Despite the growing technological advancements, it is still hard to fabricate pure nanomaterials. The existence of a disorder, such as site or bond disorder in magnetic nanoparticle systems constitutes an important role in material science, since it may induce some important macroscopic effects on the thermal and magnetic properties of these materials. For instance, magnetic properties of $\mathrm{CoFe_{2}O_{4}}$ particles with magnetically disordered surface layer, and the surface spin disorder in $\mathrm{NiFe_{2}O_{4}}$ nanoparticles have been investigated previously \cite{lin,kodama2}. In addition, spin-glass surface disorder in antiferromagnetic small particles \cite{leite}, and site dilution in surface shell of cylindrical Ising nanowire (or nanotube) systems \cite{kaneyoshi4,kaneyoshi7} have been examined, theoretically. However, to our knowledge, diluted interface and surface shell bonds in the core-shell nanoparticles have not yet been examined in detail, and the situation deserves particular attention. Therefore, aim of this study is to clarify within the framework of EFT with correlations, how the magnetism in a nanoparticle system is affected in the presence of disordered bonds in the surface shell and also in the interface between the core and shell layers of the particle. For this purpose, the paper is organized as follows: in Section \ref{formulation}, we briefly present
our model and related formulation. The results and discussions are presented in Section \ref{results}, and finally Section \ref{conclusion} is devoted to our conclusions.

\section{Formulation}\label{formulation}
\begin{figure}[!h]
\center
\includegraphics[width=6cm,height=6.5cm]{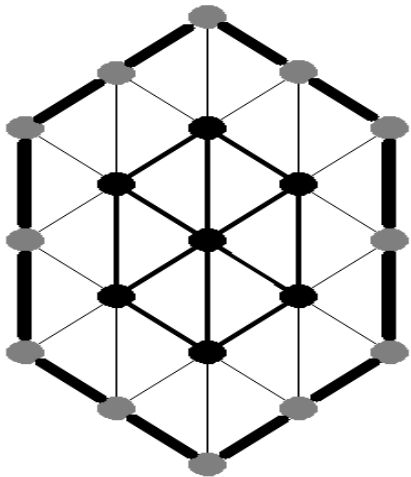}\\
\caption{Schematic representation of an infinitely long cylindrical Ising nanowire (top view). Black circles denote the magnetic atoms belonging to the core. Ferromagnetic core is surrounded by a ferromagnetic shell layer (gray circles) \cite{kaneyoshi7}.}\label{fig1}
\end{figure}
We consider an infinitely long nano-wire system composed of usual Ising spins which are located on each lattice site (Fig. \ref{fig1}). The system is composed of a ferromagnetic core which is surrounded by a ferromagnetic shell layer. At the interface, we define an exchange interaction between core and shell spins which can be either ferromagnetic or antiferromagnetic type. Hamiltonian equation describing our model can be written as
\begin{equation}\label{eq1}
\mathcal{H}=-\sum_{<ij>}J_{c}^{(i,j)}\sigma_{i}\sigma_{j}-\sum_{<kl>}J_{sh}^{(k,l)}\sigma_{k}\sigma_{l}-\sum_{<ik>}J_{int}^{(i,k)}\sigma_{i}\sigma_{k},
\end{equation}
where $\sigma=\pm1$. $J_{c}$, $J_{sh}$ and $J_{int}$ define core, shell and interface coupling parameters, respectively. Each summation in Eq. (\ref{eq1}) is carried out over the nearest neighbor spins. We assume that the nearest neighbor interactions are randomly distributed on the lattice according to the probability distribution functions
\begin{eqnarray}\label{eq2}
\nonumber
P(J_{int}^{(i,k)})&=&(1-p_{int})\delta(J_{int}^{(i,k)})+p_{int}\delta(J_{int}^{(i,k)}-J_{int}),\\
\nonumber
P(J_{c}^{(i,j)})&=&(1-p_{c})\delta(J_{c}^{(i,j)})+p_{c}\delta(J_{c}^{(i,j)}-J_{c}),\\
P(J_{sh}^{(k,l)})&=&(1-p_{sh})\delta(J_{sh}^{(k,l)})+p_{sh}\delta(J_{sh}^{(k,l)}-J_{sh}),
\end{eqnarray}
where $0<p_{c},p_{sh},p_{int}\leq 1$. These parameters denote the concentration of actively connected bonds in the core, shell and at the interface, respectively.
Following the same methodology given in Ref. \cite{kaneyoshi7}, we can get the usual coupled EFT equations for the magnetizations of core and shell layers of nanowire system as follows
\begin{eqnarray}\label{eq3}
\nonumber
m_{1}&=&[A_{1}+m_{1}B_{1}]^{4}   [A_{3}+m_{2}B_{3}]    [A_{3}+m_{3}B_{3}]^{2} \\
\nonumber
&&\times [A_{1}+m_{4}B_{1}],\\
\nonumber
m_{2}&=&[A_{3}+m_{1}B_{3}]   [A_{2}+m_{2}B_{2}]^{2}    [A_{2}+m_{3}B_{2}]^{2},\\
\nonumber
m_{3}&=&[A_{3}+m_{1}B_{3}]^{2}   [A_{2}+m_{2}B_{2}]^{2}   [A_{2}+m_{3}B_{2}]^{2},\\
m_{4}&=&[A_{1}+m_{1}B_{1}]^{6}   [A_{1}+m_{4}B_{1}]^{2},
\end{eqnarray}
where $m_{1},m_{4}$ and $m_{2},m_{3}$ terms denote the magnetizations of the core and shell sublattices, respectively. The coefficients in Eq. (\ref{eq3}) are defined as follows
\begin{eqnarray}\label{eq4}
\nonumber
A_{1}&=&\left\langle \cosh(J_{c}^{i,j})\right\rangle_{r}f(x)|_{x=0},\\
\nonumber
A_{2}&=&\left\langle \cosh(J_{sh}^{k,l})\right\rangle_{r}f(x)|_{x=0}, \\
\nonumber
A_{3}&=&\left\langle \cosh(J_{int}^{i,k})\right\rangle_{r}f(x)|_{x=0},\\
\nonumber
B_{1}&=&\left\langle \sinh(J_{c}^{i,j})\right\rangle_{r}f(x)|_{x=0},\\
\nonumber
B_{2}&=&\left\langle \sinh(J_{sh}^{k,l})\right\rangle_{r}f(x)|_{x=0},\\
B_{3}&=&\left\langle \sinh(J_{int}^{i,k})\right\rangle_{r}f(x)|_{x=0},
\end{eqnarray}
with
\begin{equation}\label{eq5}
f(x)=\tanh(\beta x),
\end{equation}
where $\beta=1/k_{B}T$, $k_{B}$ is the Boltzmann constant and $T$ is the temperature. The random configurational averages $\langle... \rangle_{r}$ in Eq. (\ref{eq4}) can be obtained by using the probability distributions given in Eq. (\ref{eq2}) and the relation $\exp(\alpha\nabla)f(x)=f(x+\alpha)$ \cite{honmura,kaneyoshi11}.

With the help of binomial expansion, Eq. (\ref{eq3}) can be written in the form
\begin{eqnarray}\label{eq6}
\nonumber
m_{1}&=&\sum_{i=0}^{4}\sum_{j=0}^{1}\sum_{k=0}^{2}\sum_{l=0}^{1}K_{1}(i,j,k,l)m_{1}^{i}m_{2}^{j}m_{3}^{k}m_{4}^{l},\\
\nonumber
m_{2}&=&\sum_{i=0}^{1}\sum_{j=0}^{2}\sum_{k=0}^{2}K_{2}(i,j,k)m_{1}^{i}m_{2}^{j}m_{3}^{k},\\
\nonumber
m_{3}&=&\sum_{i=0}^{2}\sum_{j=0}^{2}\sum_{k=0}^{2}K_{3}(i,j,k)m_{1}^{i}m_{2}^{j}m_{3}^{k},\\
m_{4}&=&\sum_{i=0}^{6}\sum_{l=0}^{2}K_{4}(i,l)m_{1}^{i}m_{4}^{l},
\end{eqnarray}
where
\begin{eqnarray}\label{eq7}
\nonumber
K_{1}(i,j,k,l)&=&\left(\begin{array}{c}4 \\ i \\ \end{array}\right) \left(\begin{array}{c}1 \\ j \\ \end{array}\right)  \left(\begin{array}{c}2 \\ k \\ \end{array}\right)  \left(\begin{array}{c}1 \\ l \\ \end{array}\right)  \\
\nonumber
&&\times A_{1}^{5-i-l}A_{3}^{3-j-k}B_{1}^{i+l}B_{3}^{j+k},\\
\nonumber
K_{2}(i,j,k)&=& \left(\begin{array}{c}1 \\ i \\ \end{array}\right) \left(\begin{array}{c}2 \\ j \\ \end{array}\right) \left(\begin{array}{c}2 \\ k \\ \end{array}\right)A_{2}^{4-j-k}A_{3}^{1-i}B_{2}^{j+k}B_{3}^{i},\\
\nonumber
K_{3}(i,j,k)&=&\left(\begin{array}{c}2 \\ i \\ \end{array}\right) \left(\begin{array}{c}2 \\ j \\ \end{array}\right) \left(\begin{array}{c}2 \\ k \\ \end{array}\right) A_{2}^{4-j-k}A_{3}^{2-i}B_{2}^{j+k}B_{3}^{i},\\
K_{4}(i,l)&=&\left(\begin{array}{c}6 \\ i \\ \end{array}\right) \left(\begin{array}{c}2 \\ l \\ \end{array}\right) A_{1}^{8-i-l}B_{1}^{i+l}.
\end{eqnarray}
Once the coefficients given in Eq. (\ref{eq7}) are calculated, we can construct a system of nonlinear coupled equations from Eq. (\ref{eq6}) and obtain the magnetizations $m_{i}, \quad i=1,2,3,4$. Magnetizations of the core $(m_{c})$, shell $(m_{sh})$, and the total magnetization $(m_{T})$ of the system are defined as
\begin{equation}\label{eq8}
m_{c}=\frac{1}{7}(m_{1}+m_{4}),\quad m_{sh}=\frac{1}{12}(6m_{2}+6m_{3}),\quad m_{T}=\frac{1}{19}(7m_{c}+12m_{sh}).
\end{equation}

In the vicinity of the transition temperature, we have  $m_{i}(T\rightarrow T_{c})\simeq0$. Hence, in order to obtain the critical temperature we may linearize Eq. (\ref{eq6}), i.e.,
\begin{equation}\label{eq9}
Am=0,
\end{equation}
where
{\small
\begin{eqnarray}
\nonumber
A&=&
\end{eqnarray}
\begin{eqnarray}\label{eq10}
\nonumber
\left(
    \begin{array}{cccc}
      K_{1}(1,0,0,0)-1 & K_{1}(0,1,0,0) & K_{1}(0,0,1,0) & K_{1}(0,0,0,1) \\
      K_{2}(1,0,0) & K_{2}(0,1,0)-1 & K_{2}(0,0,1) & 0 \\
      K_{3}(1,0,0) & K_{3}(0,1,0) & K_{3}(0,0,1)-1 & 0 \\
      K_{4}(1,0) & 0 & 0 & K_{4}(0,1)-1 \\
    \end{array}
  \right),
\end{eqnarray}}
and
\begin{eqnarray}\label{eq11}
m=\left(
    \begin{array}{c}
      m_{1} \\
      m_{2} \\
      m_{3} \\
      m_{4} \\
    \end{array}
  \right).
\end{eqnarray}
For the selected values of the system parameters, the transition temperature of the system can be obtained from $\mathrm{det}(A)=0$. Finally,
we should note that, as is discussed in Ref. \cite{kaneyoshi7}, Eq. (\ref{eq10}) is invariant under
the transformation $J_{int}\rightarrow -J_{int}$ which means that the phase diagrams are independent of the sign of $J_{int}$.

\section{Results and Discussion}\label{results}
In this section, we will discuss how the disordered shell and interface bonds of the particle affect the phase diagrams of the system.
In the following discussions, we consider a pure core layer with $p_{c}=1.0$, and also we fix the exchange interaction of the
core spins as $J_{c}=1.0$. We will also consider both ferromagnetic $(J_{int}>0)$ and antiferromagnetic $(J_{int}<0)$ interface coupling properties of the system.
As is stated above, phase diagrams corresponding to ferromagnetic interface coupling are also valid for the system with antiferromagnetic interface coupling. Critical
temperature of the system at which a phase transition occurs does not change since Eq. (\ref{eq10}) is invariant under the transformation $J_{int}\rightarrow -J_{int}$. However, in contrast to
the ferromagnetic case, the nanowire system with a negative interface coupling may exhibit a novel phenomenon, namely a compensation point which does not occur in the ferromagnetic case.
Compensation point is of particular interest, since this phenomenon has important technological applications in magneto-optical recording devices. In addition,  nanoscaled magnets
such as nanowires, nanotubes, etc. are currently considered as promising candidates due to their potential utilization as ultra-high density recording media.

\subsection{Nanowire with ferromagnetic interface coupling $\mathrm{(J_{int}>0)}$:}\label{results_a}
\begin{figure*}[!h]
\center
\includegraphics[width=8cm]{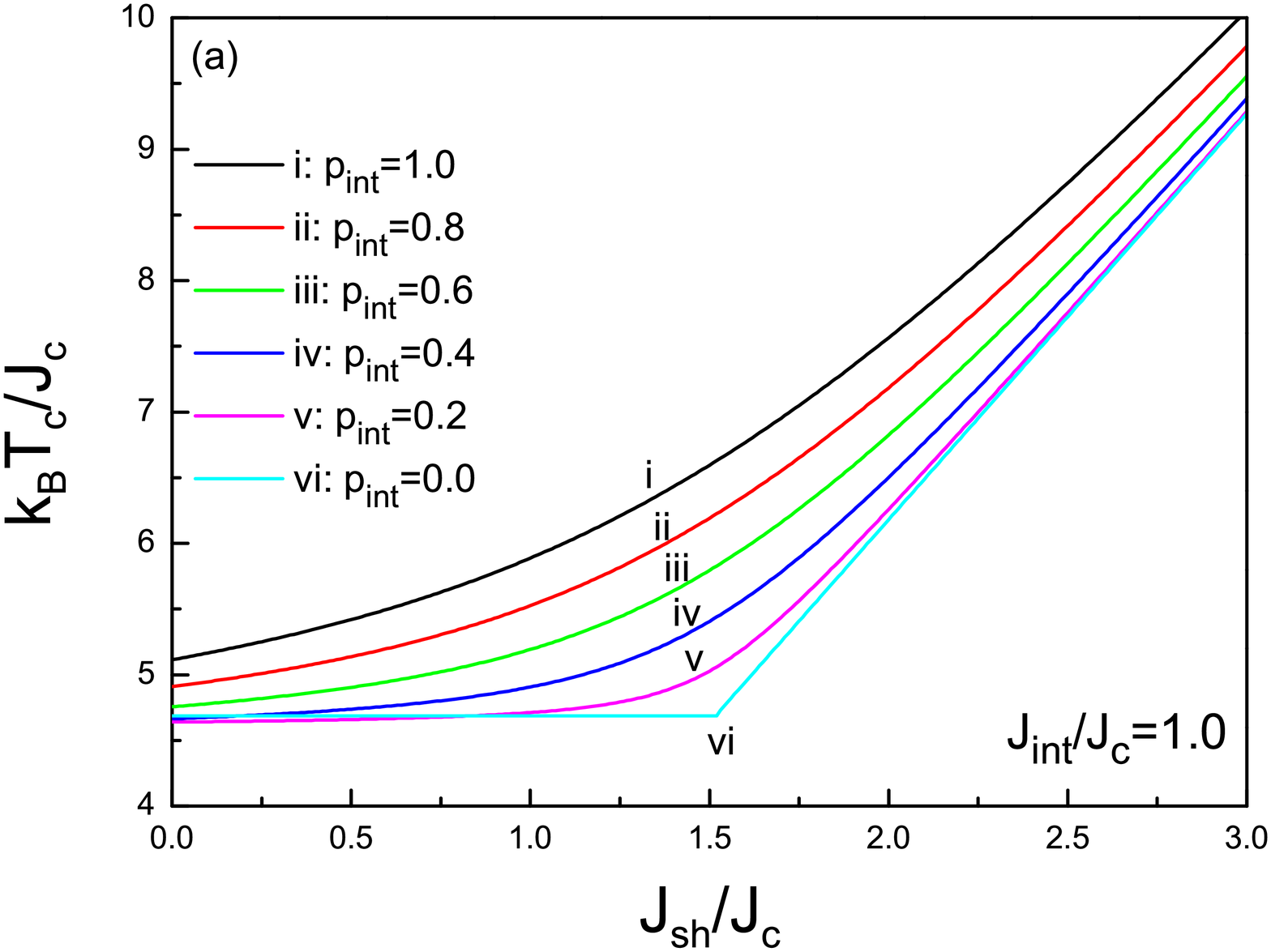}
\includegraphics[width=8cm]{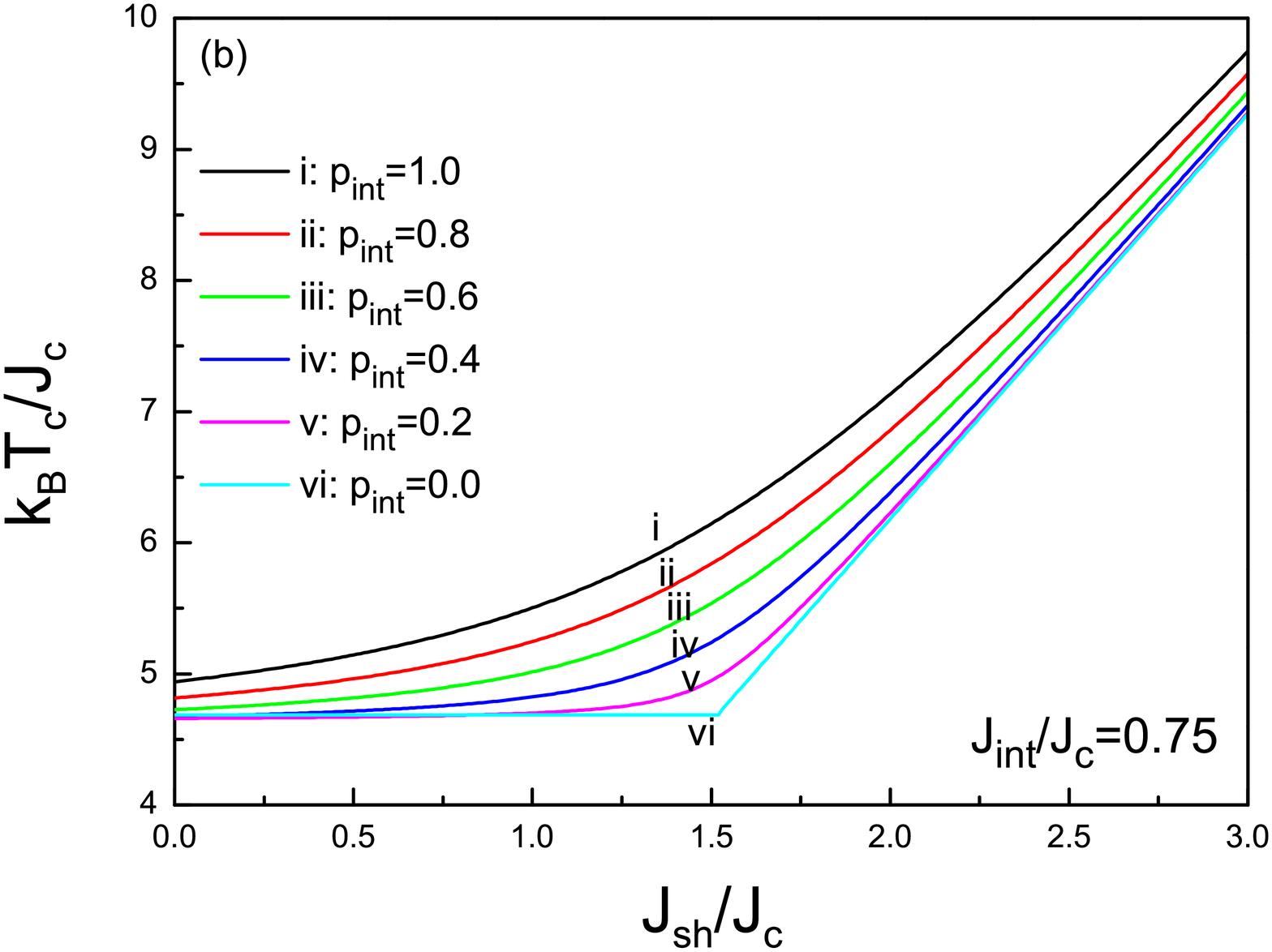}\\
\includegraphics[width=8cm]{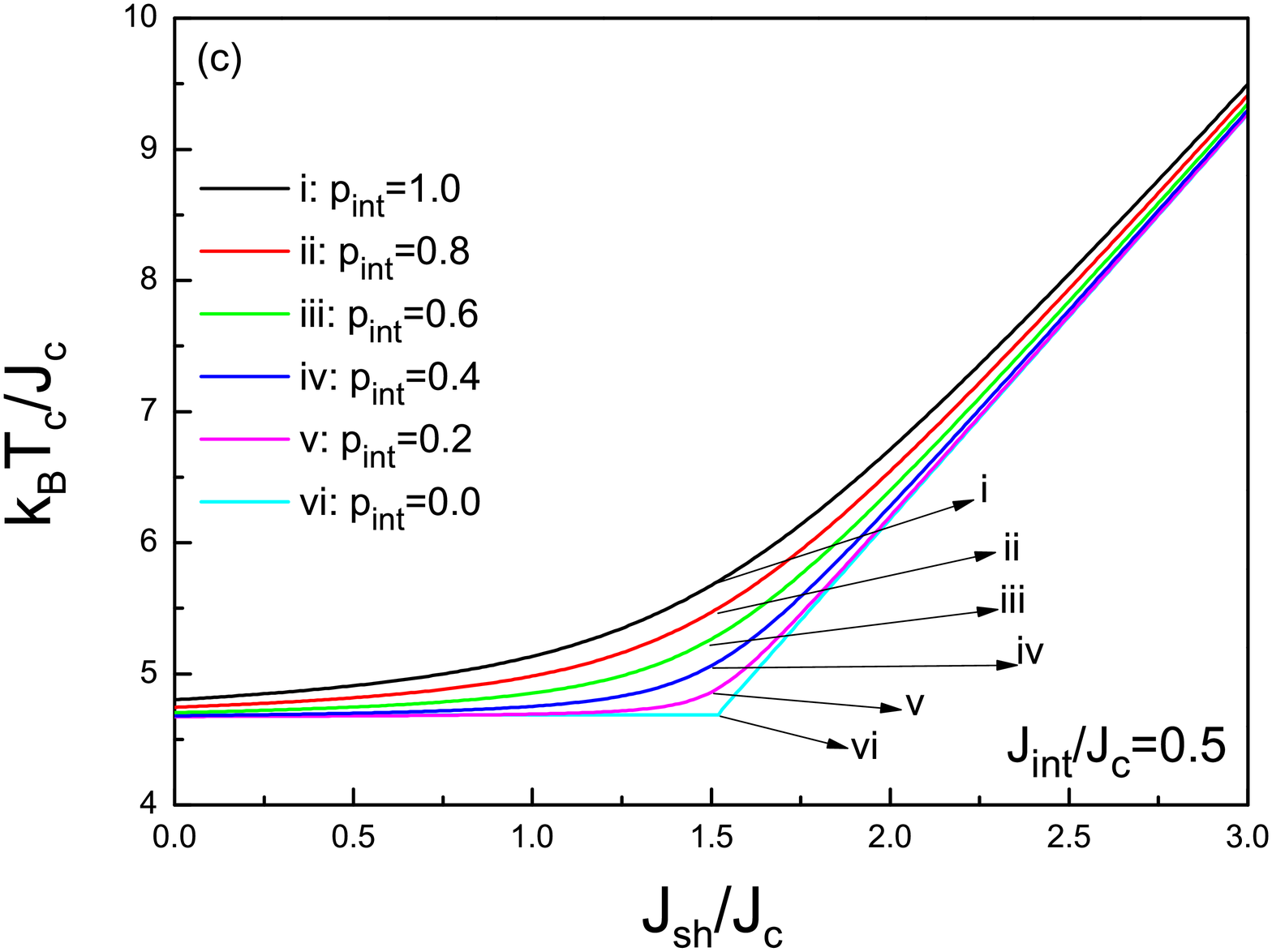}
\includegraphics[width=8cm]{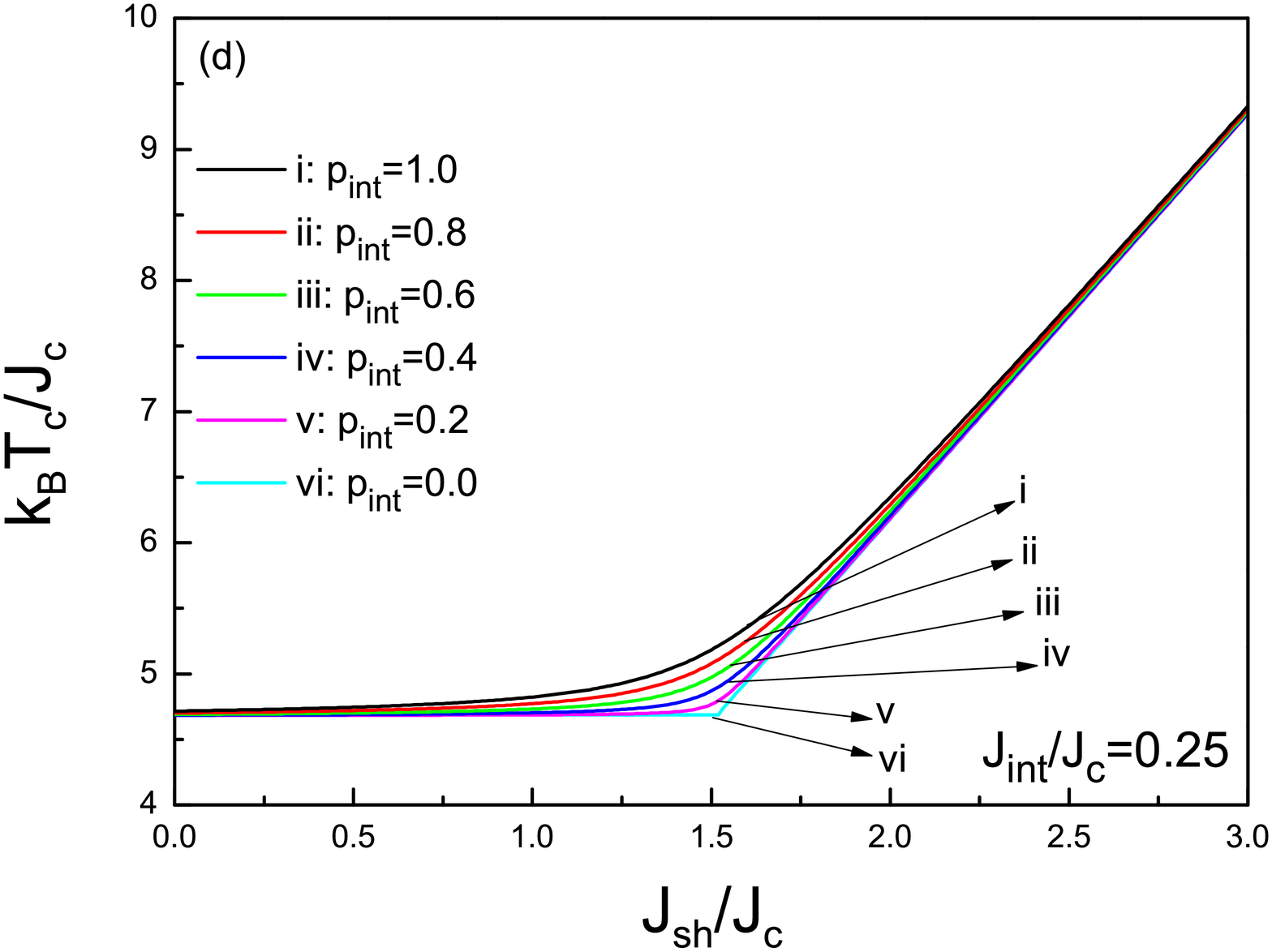}\\
\caption{Phase diagrams of the ferromagnetic nanowire system in a $(k_{B}T_{c}/J_{c}-J_{sh}/J_{c})$ plane with four selected values of interface coupling $J_{int}/J_{c}$:
(a) $J_{int}/J_{c}=1.0$, (b) $J_{int}/J_{c}=0.75$, (c) $J_{int}/J_{c}=0.5$, (d) $J_{int}/J_{c}=0.25$ with various interface concentrations. In all figures, $p_{c}=1.0$, $p_{sh}=1.0$}\label{fig2}
\end{figure*}
\begin{figure}[!h]
\center
\includegraphics[width=8cm]{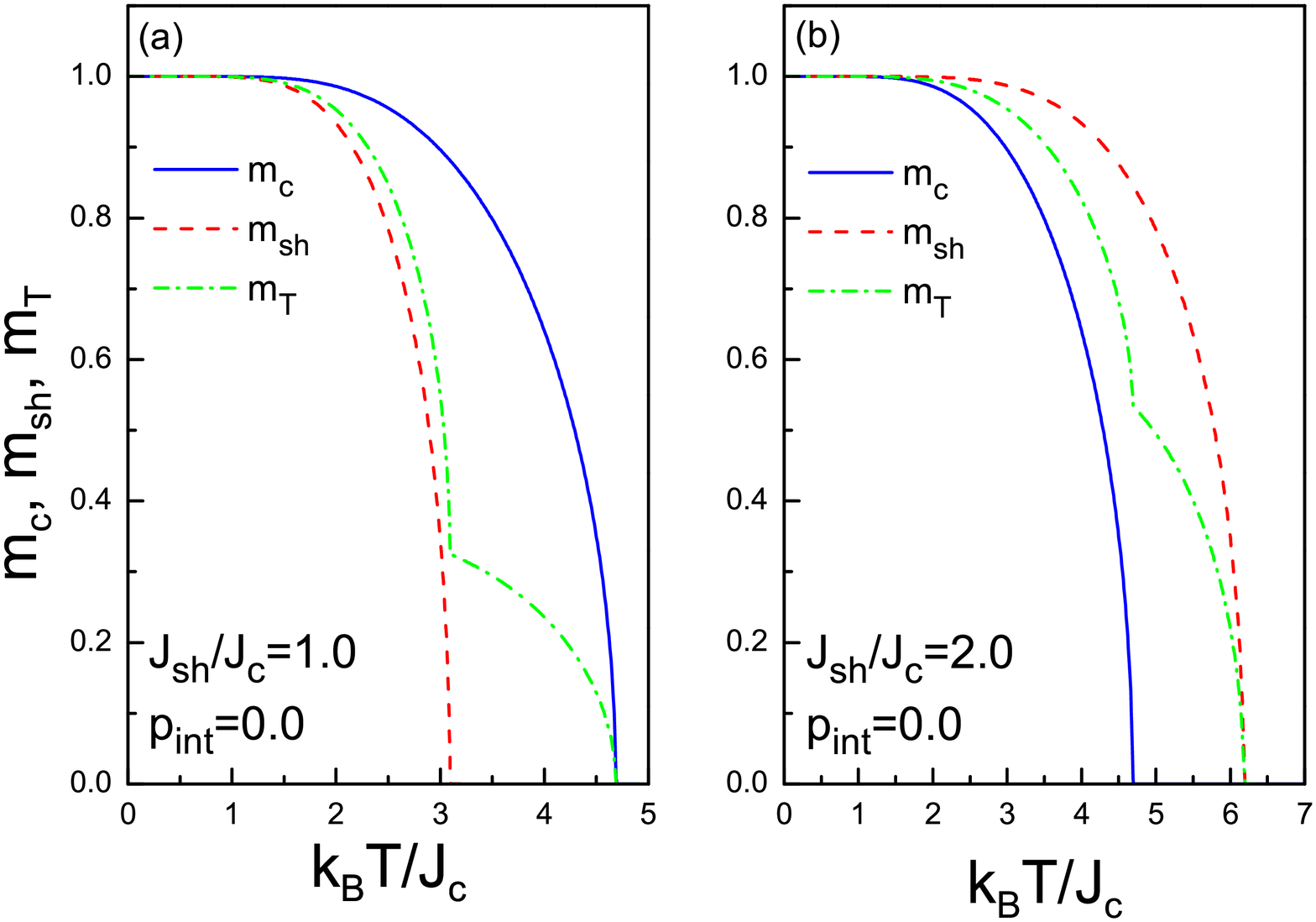}\\
\caption{Core (solid line), shell (dashed line) and total magnetization (dashed-dotted line) curves as functions of the temperature in the nanowire system, corresponding to the phase diagrams
in Fig. \ref{fig2}a with $J_{int}/J_{c}=1.0$, $p_{c}=1.0$, $p_{sh}=1.0$}\label{fig3}
\end{figure}
\begin{figure}[!h]
\center
\includegraphics[width=8cm]{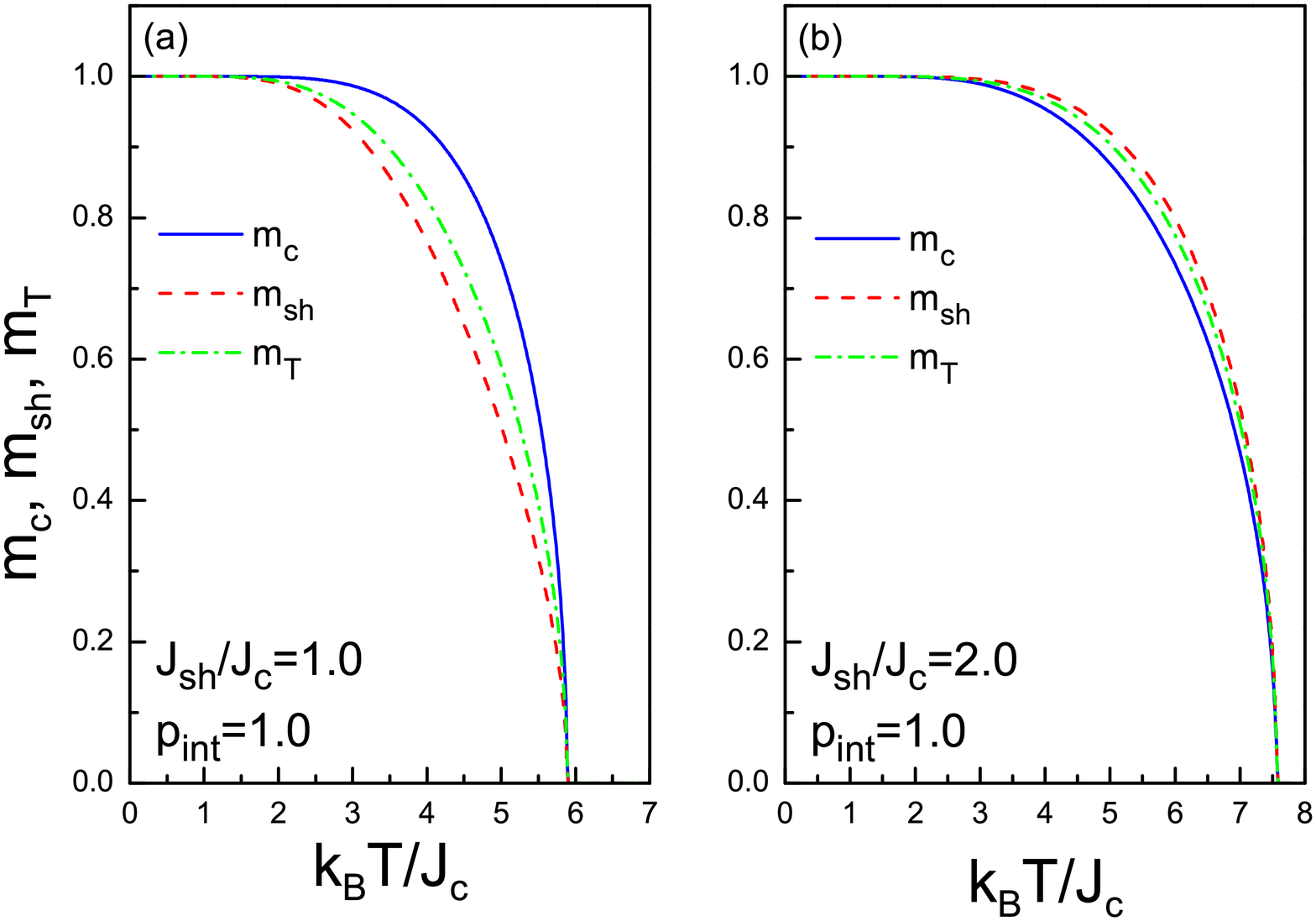}\\
\caption{Core (solid line), shell (dashed line) and total magnetization (dashed-dotted line) curves as functions of the temperature in the nanowire system, corresponding to the phase diagrams
in Fig. \ref{fig2}a with $J_{int}/J_{c}=1.0$ $p_{c}=1.0$, $p_{sh}=1.0$}\label{fig4}
\end{figure}

In Fig. \ref{fig2}, we present the phase diagrams of the system in a $(k_{B}T_{c}/J_{c}-J_{sh}/J_{c})$ plane with four selected values of
interface coupling $J_{int}/J_{c}$ with different concentrations, namely with various $p_{int}$ values. The first conspicuous observation
in Fig. \ref{fig2} is that increasing ferromagnetic shell coupling $J_{sh}/J_{c}$ affects the critical temperature of the system in two
apparent characteristics. For $p_{int}>0$, critical temperature of the system gradually increases as the strength of the ferromagnetic
shell coupling increases whereas for $p_{int}=0$, critical temperature of the system remains constant $(T_{c}=4.6854)$ up to a threshold
value of $J_{sh}/J_{c}$ which is independent of $J_{int}/J_{c}$, and the system exhibits similar phase transition characteristics observed
in the semi-infinite ferromagnets \cite{kaneyoshi_final}. Moreover, critical temperature of the system decreases as diluting or weakening
the interface bonds of the system. In Figs. \ref{fig3} and \ref{fig4}, we plot the core, shell and total magnetizations of the particle
corresponding to the phase diagrams in Fig. \ref{fig2}a. As seen in Fig. \ref{fig3}a, when all of the interface bonds are broken
$(\mathrm{i.e.} p_{int}=0 )$, core and shell layers of the particle become independent of each other and they have different
transition temperatures. In this case, although the strength of the exchange interactions are equal to each other $(J_{sh}/J_{c}=1)$,
transition temperature of the shell layer is lower than that of the particle core, since the shell and core layers of the particle have
different coordination numbers. Horizontal line $(T_{c}=4.6854)$ in Fig. \ref{fig2} is the transition temperature of core layer of the particle,
since the contribution of the particle shell to the transition temperature is very weak. When the ferromagnetic shell coupling exceeds the threshold
value, such as for $J_{sh}/J_{c}=2.0$ in Fig. \ref{fig3}b, the effect of ferromagnetic shell interactions becomes dominant and the transition
temperature of the shell layer becomes greater than that of the core, and hence, the linearly increasing part of critical line originates
from the transition temperature of the particle shell. On the other hand, the core, shell, and the total magnetization curves of the nanowire
system with pure bonds are depicted in Fig. \ref{fig4}. In this figure, all of the ferromagnetic exchange couplings are active, hence $p_{c}=p_{sh}=p_{int}=1.0$.
In Figs. \ref{fig4}a and \ref{fig4}b, we plot the magnetization curves for $J_{int}/J_{c}=1.0$ with $J_{sh}/J_{c}=1.0$, and $J_{sh}/J_{c}=2.0$, respectively.
In this case, we can clearly see in Fig. \ref{fig4} that, for large values of interface coupling such as $J_{int}/J_{c}=1.0$, the transition
temperature of the nanowire system does not exhibit any horizontal line, since both the core and shell layers of the particle have the same
transition temperature values, even the exchange coupling constants are different from each other (see Fig. \ref{fig4}b). Similar results have been observed previously in Refs.\cite{kaneyoshi7,yuksel}. However, we show that the same phenomena can be observed not only for weak interface
coupling, but also in the presence of diluted interface bonds.

\begin{figure*}[!h]
\center
\includegraphics[width=8cm]{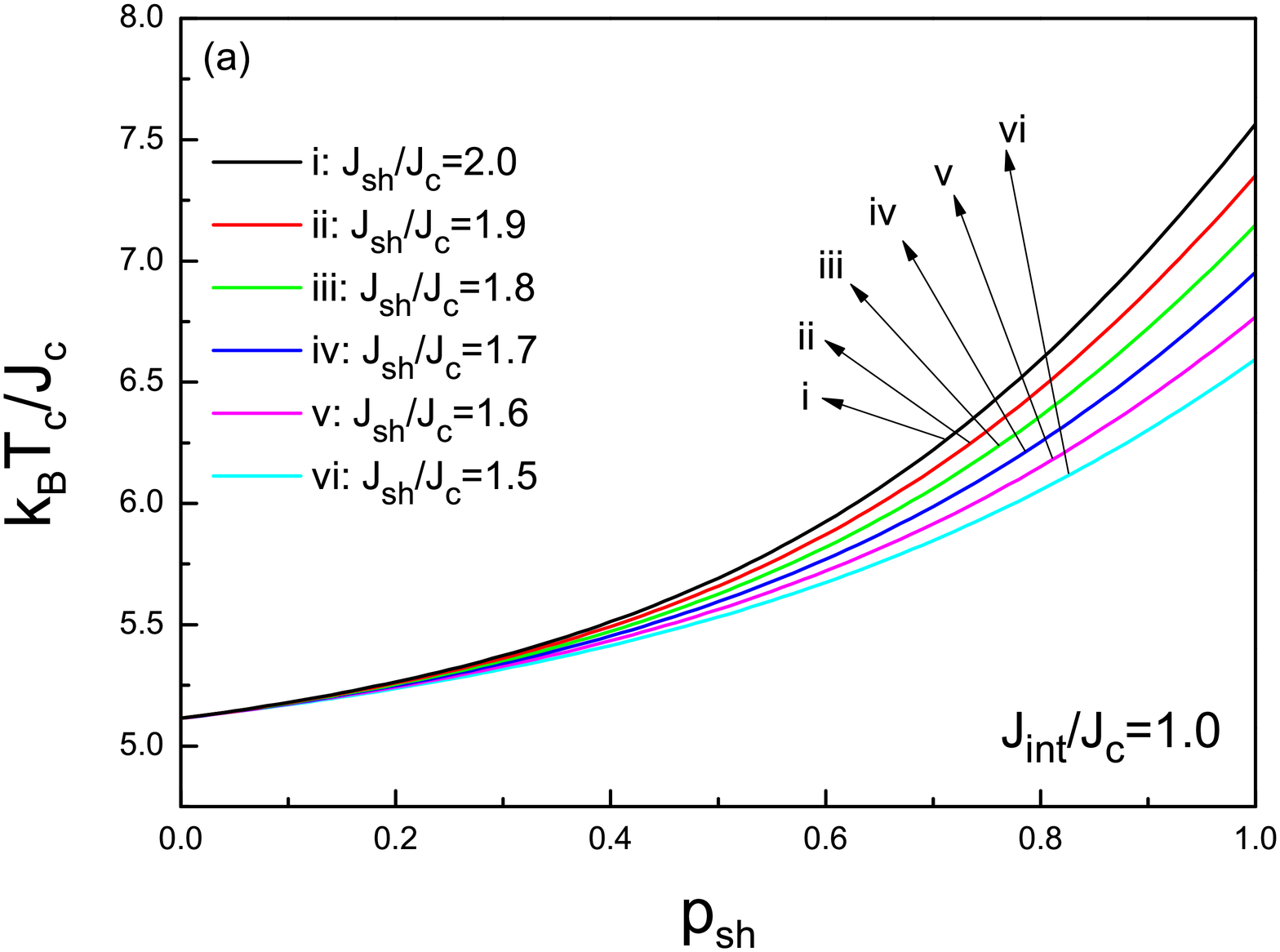}
\includegraphics[width=8cm]{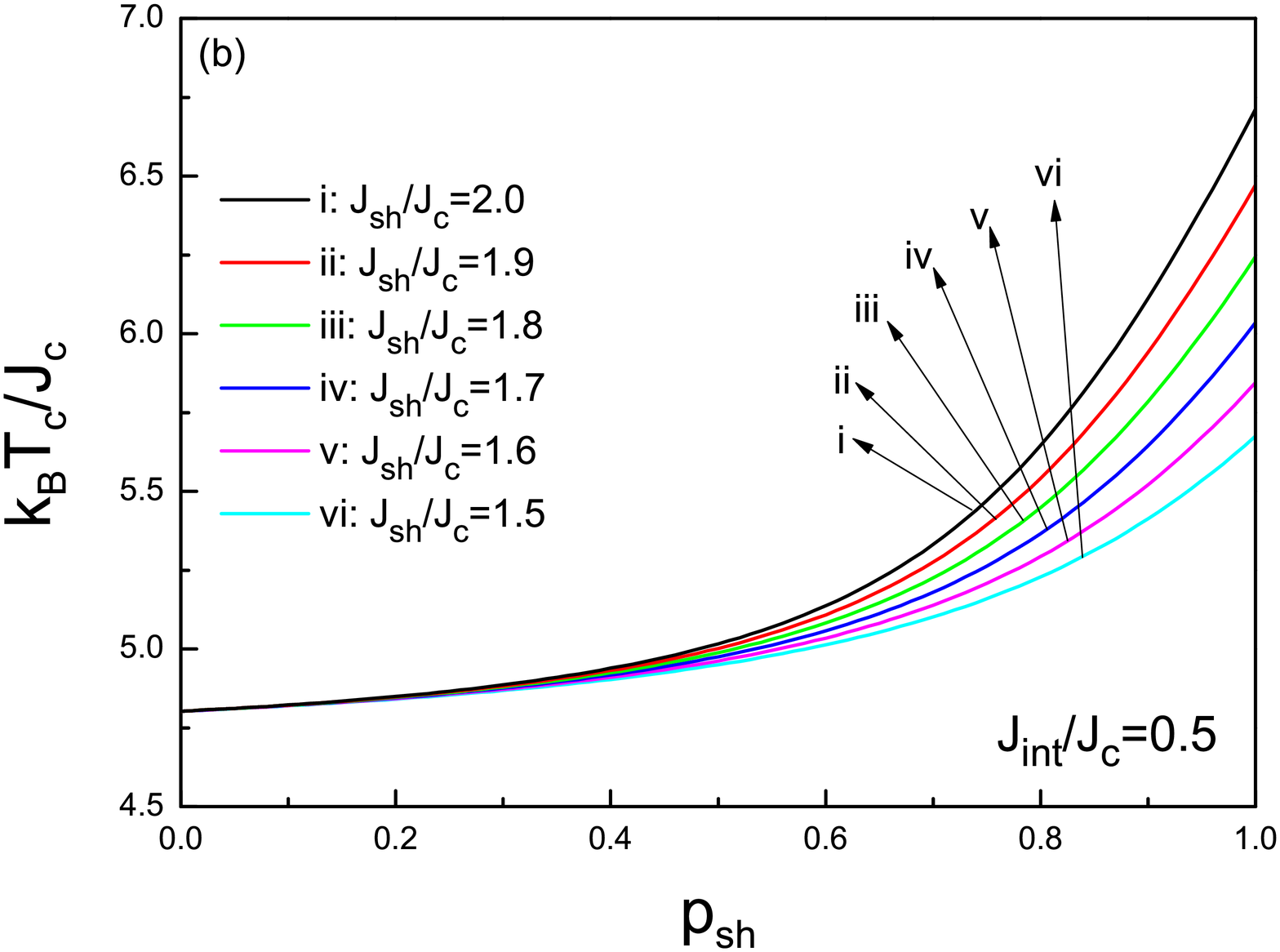}\\
\includegraphics[width=8cm]{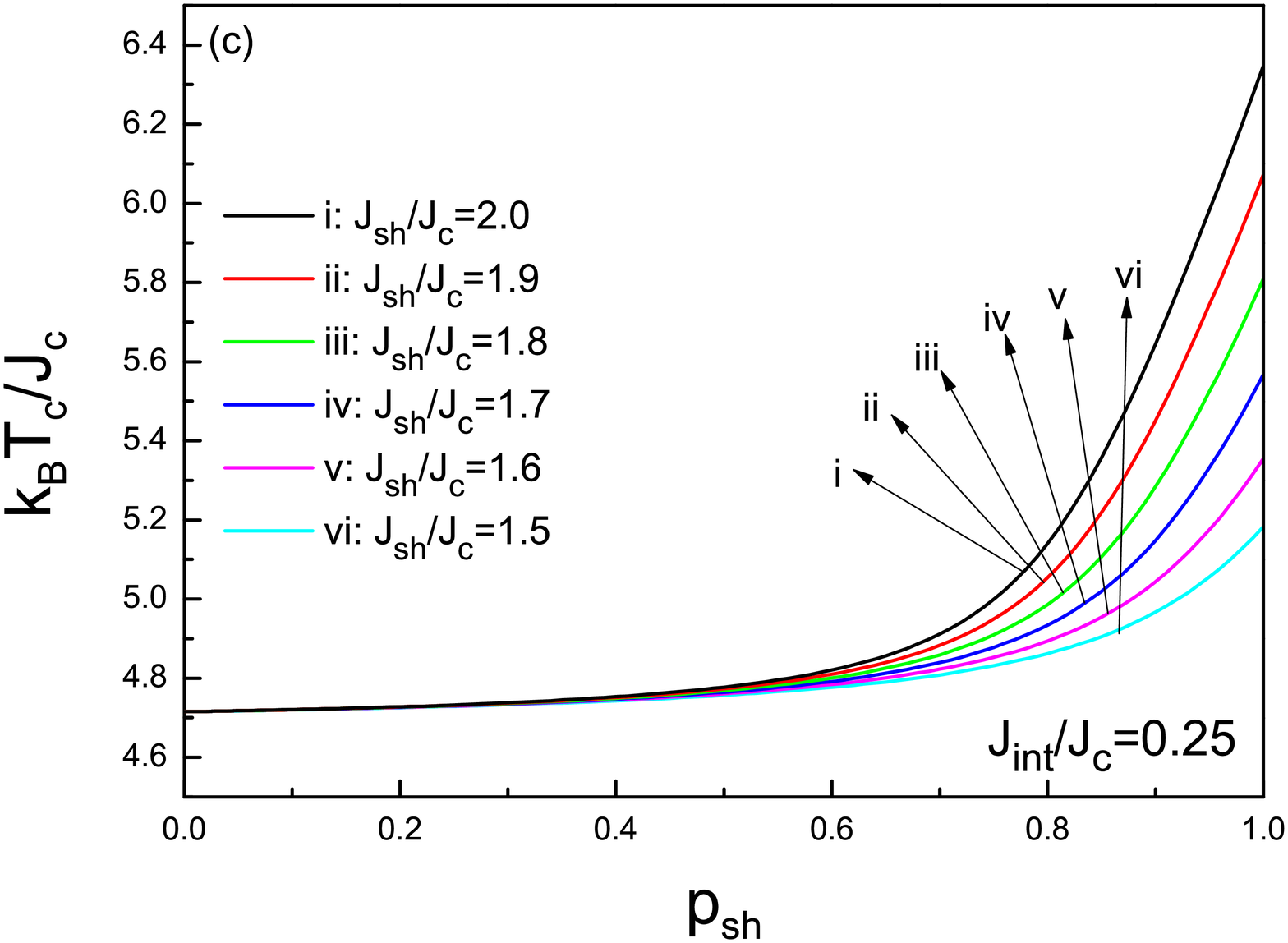}
\includegraphics[width=8cm]{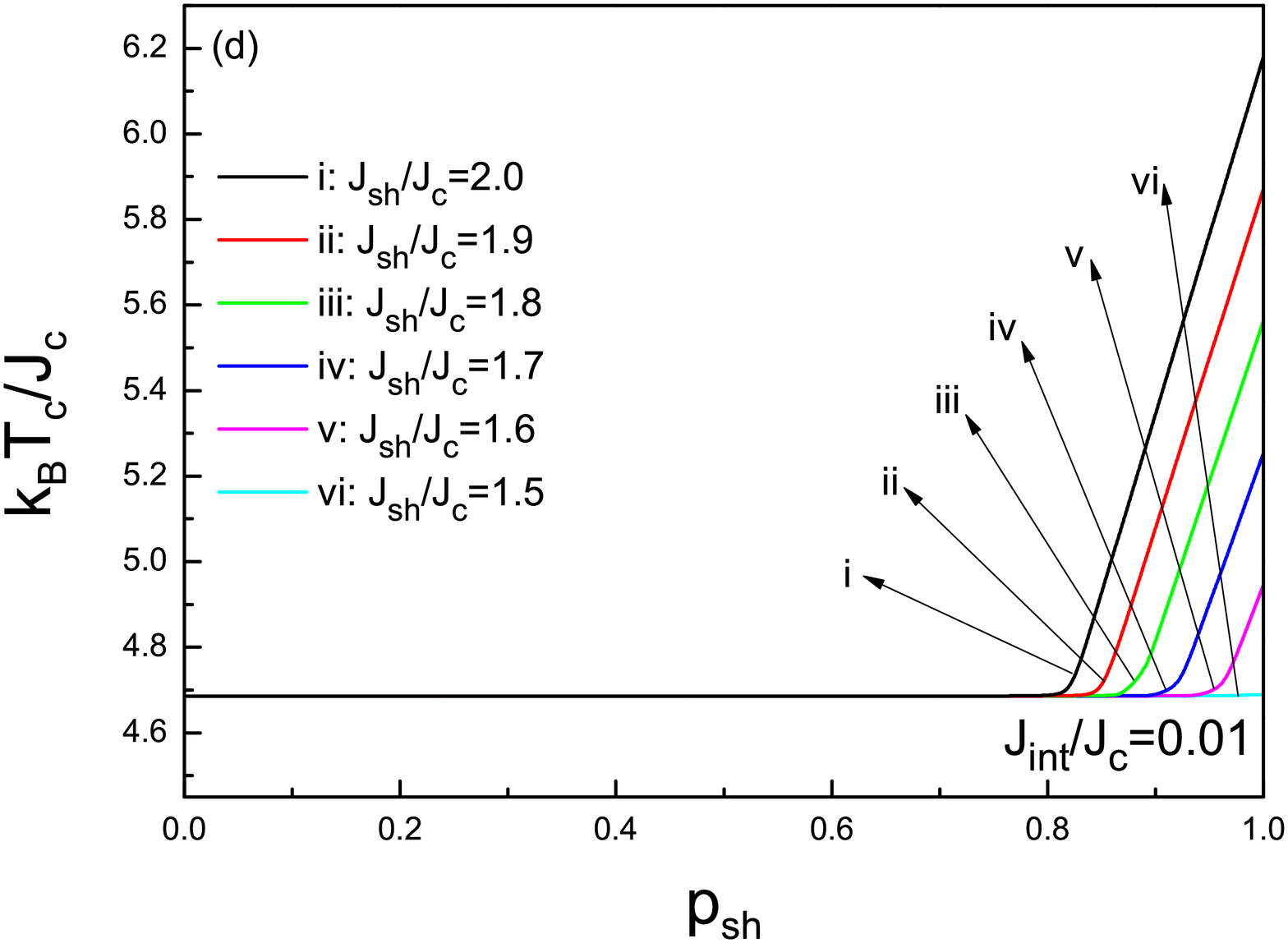}\\
\caption{Phase diagrams of the ferromagnetic nanowire system in a $(k_{B}T_{c}/J_{c}-p_{sh})$ plane with four selected values of interface coupling
$J_{int}/J_{c}$: (a) $J_{int}/J_{c}=1.0$, (b) $J_{int}/J_{c}=0.5$, (c) $J_{int}/J_{c}=0.25$, (d) $J_{int}/J_{c}=0.01$ for various ferromagnetic shell coupling strengths $J_{sh}/J_{c}$
with $p_{int}=1.0$, $p_{c}=1.0$}\label{fig5}
\end{figure*}
\begin{figure}[!h]
\center
\includegraphics[width=8cm]{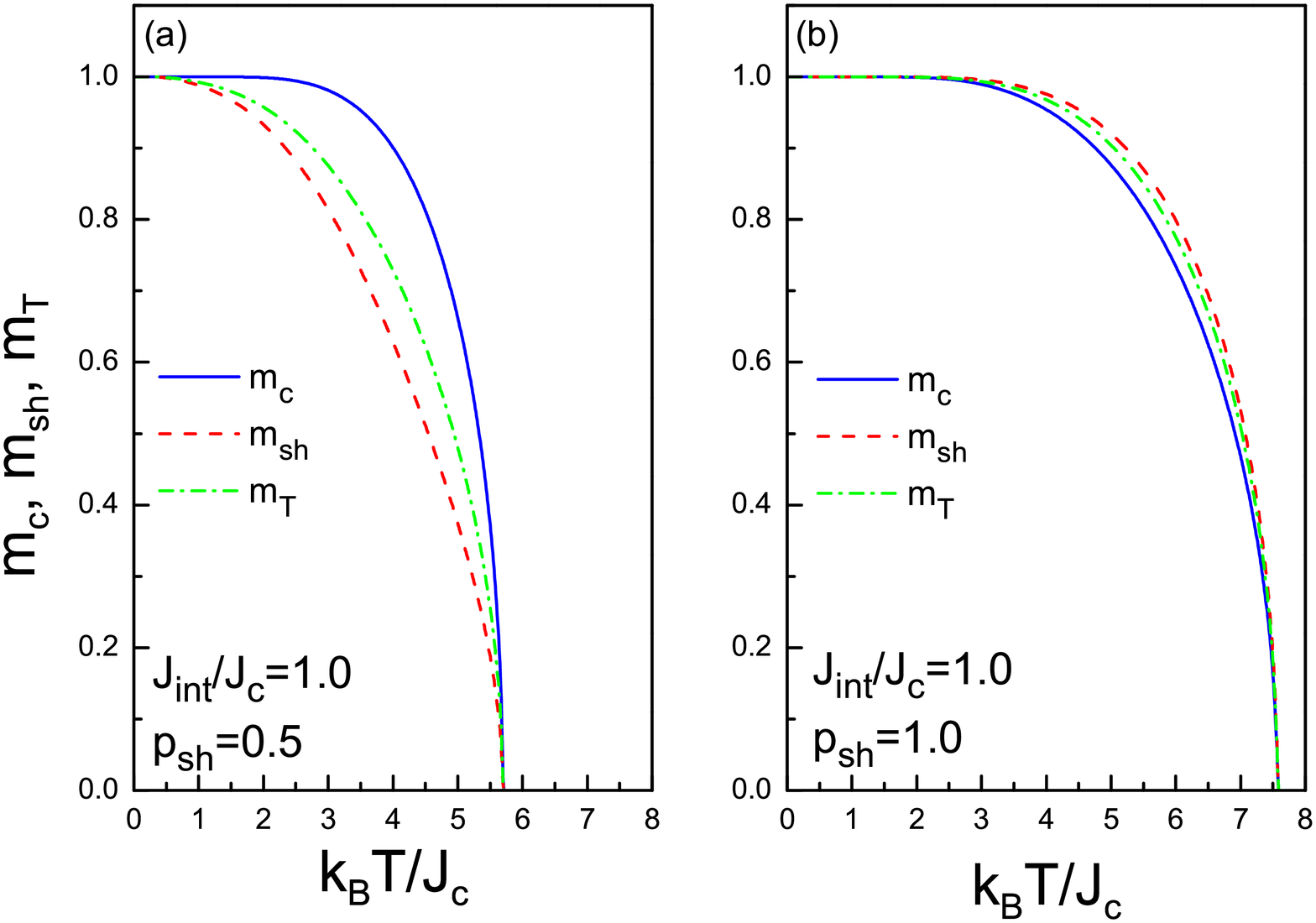}\\
\caption{Core (solid line), shell (dashed line) and total magnetization (dashed-dotted line) curves as functions of the temperature in the nanowire system, corresponding to the phase diagrams
in Fig. \ref{fig5}a with $J_{sh}/J_{c}=2.0$, $p_{int}=1.0$, $p_{c}=1.0$}\label{fig6}
\end{figure}
\begin{figure}[!h]
\center
\includegraphics[width=8cm]{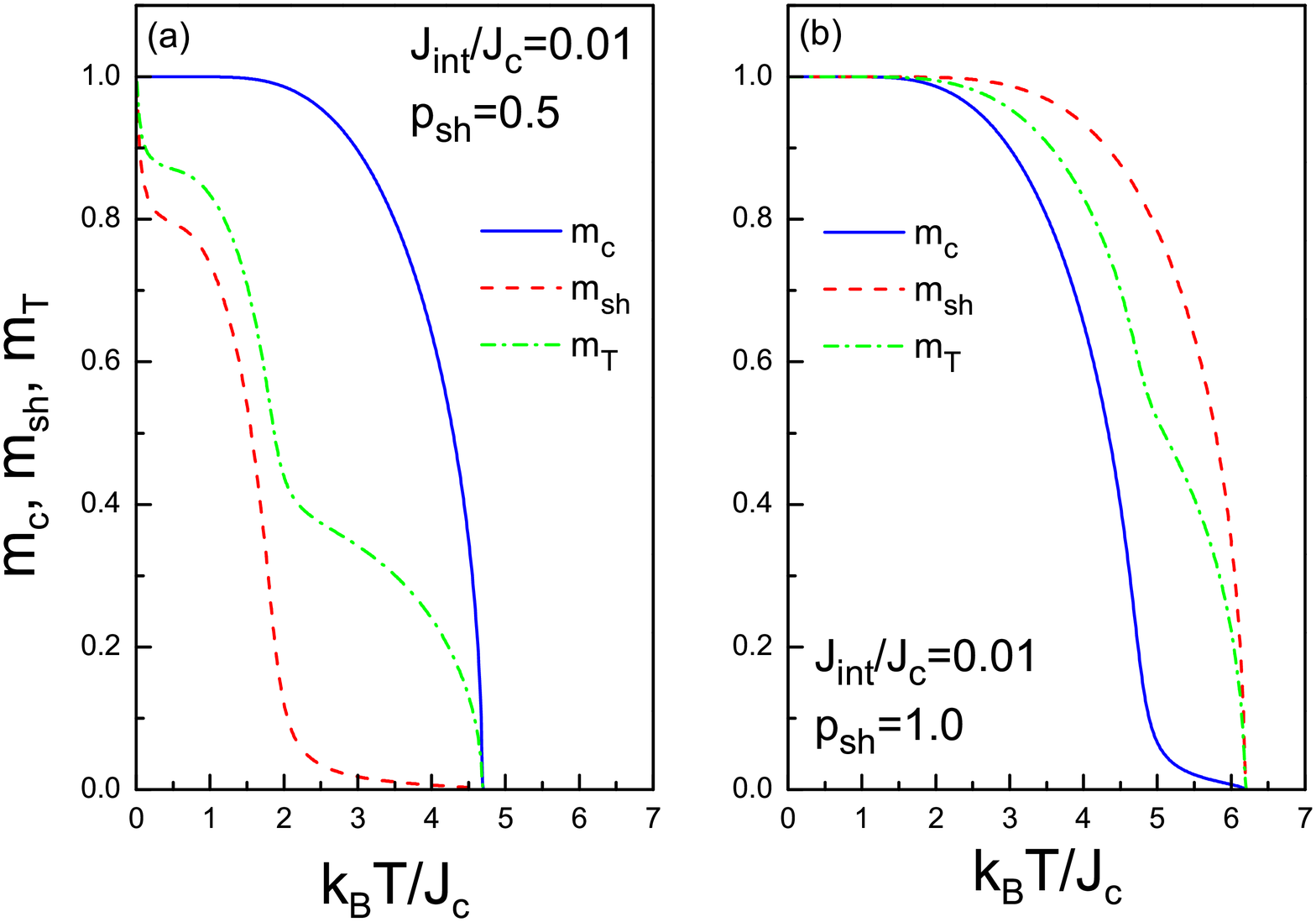}\\
\caption{Core (solid line), shell (dashed line) and total magnetization (dashed-dotted line) curves as functions of the temperature in the nanowire system, corresponding to the phase diagrams
in Fig. \ref{fig5}d with $J_{sh}/J_{c}=2.0$, $p_{int}=1.0$, $p_{c}=1.0$}\label{fig7}
\end{figure}

In Fig. \ref{fig5}, in order to investigate the effect of the bond dilution process in the surface shell on the phase transition properties of the nanowire system,
we depict the phase diagrams in $(k_{B}T_{c}/J_{c}-p_{sh})$ plane with four selected values of ferromagnetic interface coupling, namely $J_{int}/J_{c}=1.0,0.5,0.25$ and $0.01$  for
some selected values of  ferromagnetic shell coupling values $J_{sh}/J_{c}$. In this figure, it is clear that as the ferromagnetic interface and shell couplings are
strengthened then the transition temperature of the system increases and ferromagnetic phase region in the phase diagrams in $(k_{B}T_{c}/J_{c}-p_{sh})$ plane gets wider, as expected.
In addition, for sufficiently weak interface bonds, such as for $J_{int}/J_{c}=0.01$ in Fig. \ref{fig5}d, transition temperature of the
system does not change until a threshold concentration in the surface bonds $p_{sh}^{*}$ is reached. This critical value depends on the value of ferromagnetic shell coupling $J_{sh}/J_{c}$.
Namely, as $J_{sh}/J_{c}$ increases then $p_{sh}^{*}$ decreases. For $p_{sh}=0$, and in the limit $J_{int}\rightarrow0$ or $p_{int}\rightarrow0$, all transition curves in Fig. \ref{fig5},
converge to the transition temperature of the core. This result agrees well with that found in Ref. \cite{kaneyoshi7} where the magnetic surface shell sites are diluted instead of the bonds.
Thermal variation of core, shell and total magnetization curves corresponding to the phase diagrams in Figs. \ref{fig5}a and \ref{fig5}d are plotted in Figs. \ref{fig6} and \ref{fig7},
respectively. As seen in Figs. \ref{fig6}a and \ref{fig6}b, as the concentration of ferromagnetic shell coupling increases
then the transition temperature of the system also increases. Moreover, according to Figs. \ref{fig6}a and \ref{fig6}b, when the interface coupling is strong enough, such as in
the case $p_{int}=1.0$ and $J_{int}/J_{c}=1.0$, both the particle core and shell layers have the same transition temperature. Hence, as shown in Figs. \ref{fig5}a-\ref{fig5}c, transition
temperature of the system gradually increases with increasing $p_{sh}$. On the other hand, as shown in Fig. \ref{fig7}, for pure, but weak ferromagnetic interface interactions, core magnetization
exhibits a larger curve in comparison with the shell magnetization curve when the concentration of the surface shell layer $p_{sh}$ is low (Fig.\ref{fig7}a). However, after some
threshold value $p_{sh}^{*}$, shell magnetization curve gets wider than the core magnetization (Fig.\ref{fig7}b). As shown in Fig. \ref{fig5}d, the threshold value of
the surface bond concentration $p_{sh}^{*}$ depends on the strength of the ferromagnetic surface coupling $J_{sh}/J_{c}$. The results obtained in Figs. \ref{fig3}, \ref{fig4}
and Figs. \ref{fig6} and \ref{fig7} are consistent with the phase diagrams presented in Figs. \ref{fig2} and \ref{fig5}, respectively.

\subsection{Nanowire with antiferromagnetic interface coupling $\mathrm{(J_{int}<0:)}$}\label{results_b}
\begin{figure}[!h]
\center
\includegraphics[width=8cm]{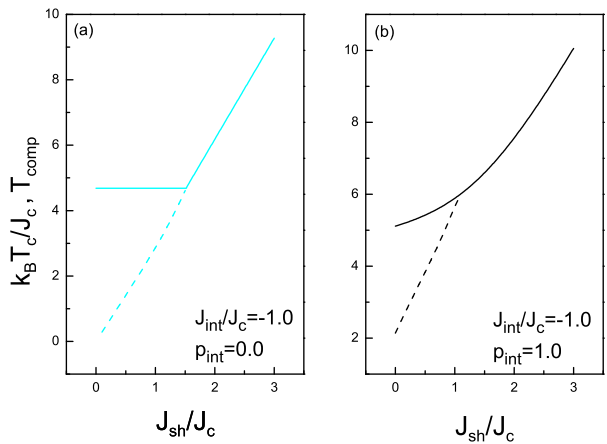}\\
\caption{Variation of critical and compensation temperatures of the antiferromagnetic nanowire system as a function of ferromagnetic shell coupling $J_{sh}/J_{c}$ for (a) $p_{int}=0.0$, (b) $p_{int}=1.0$. In both panels, we set $J_{int}/J_{c}=-1.0$, $p_{c}=1.0$ and $p_{sh}=1.0$. Solid (dashed) line represent the transition temperature (compensation point) of the system.}\label{fig8}
\end{figure}
\begin{figure}[!h]
\center
\includegraphics[width=8cm]{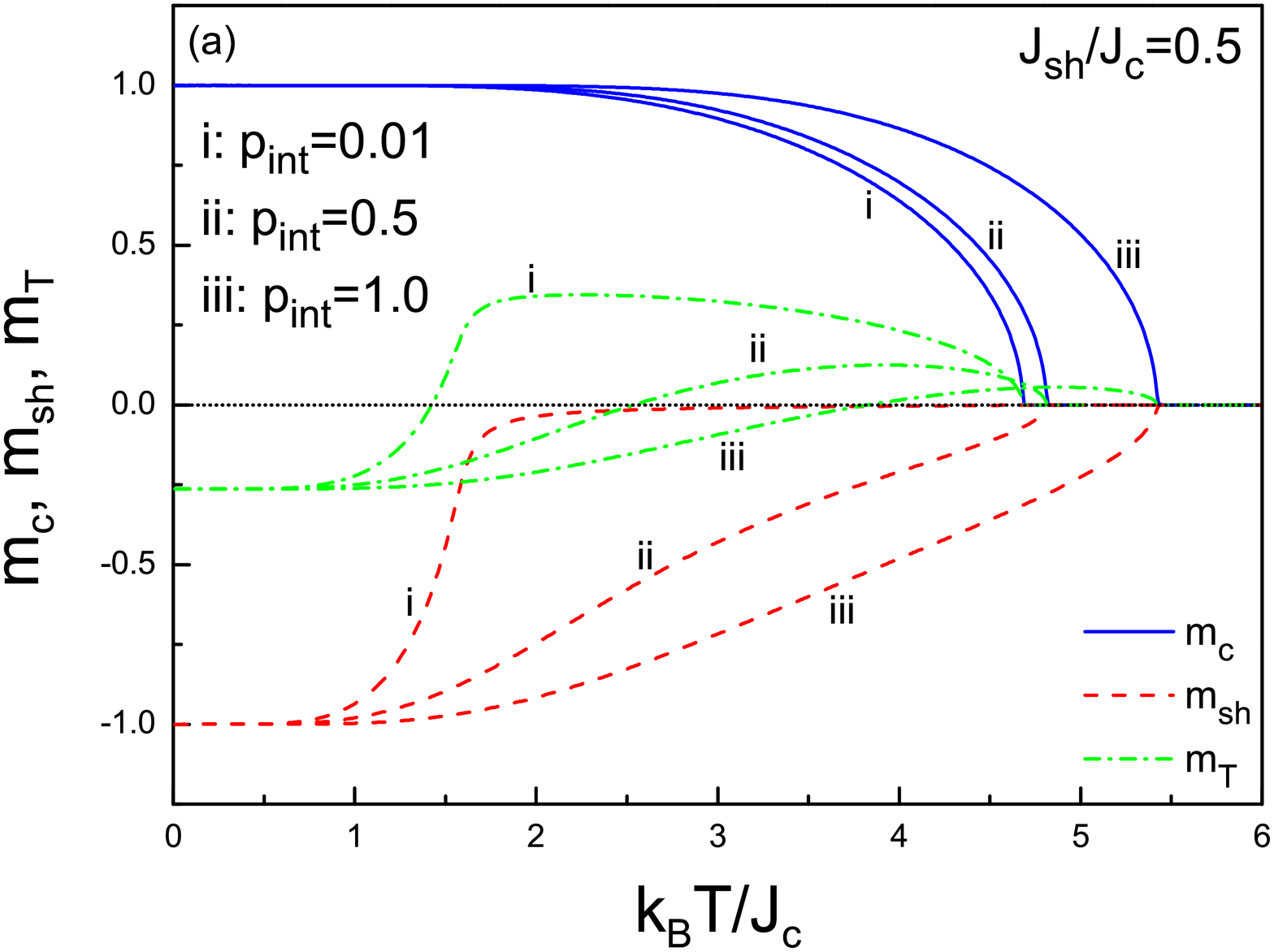}
\includegraphics[width=8cm]{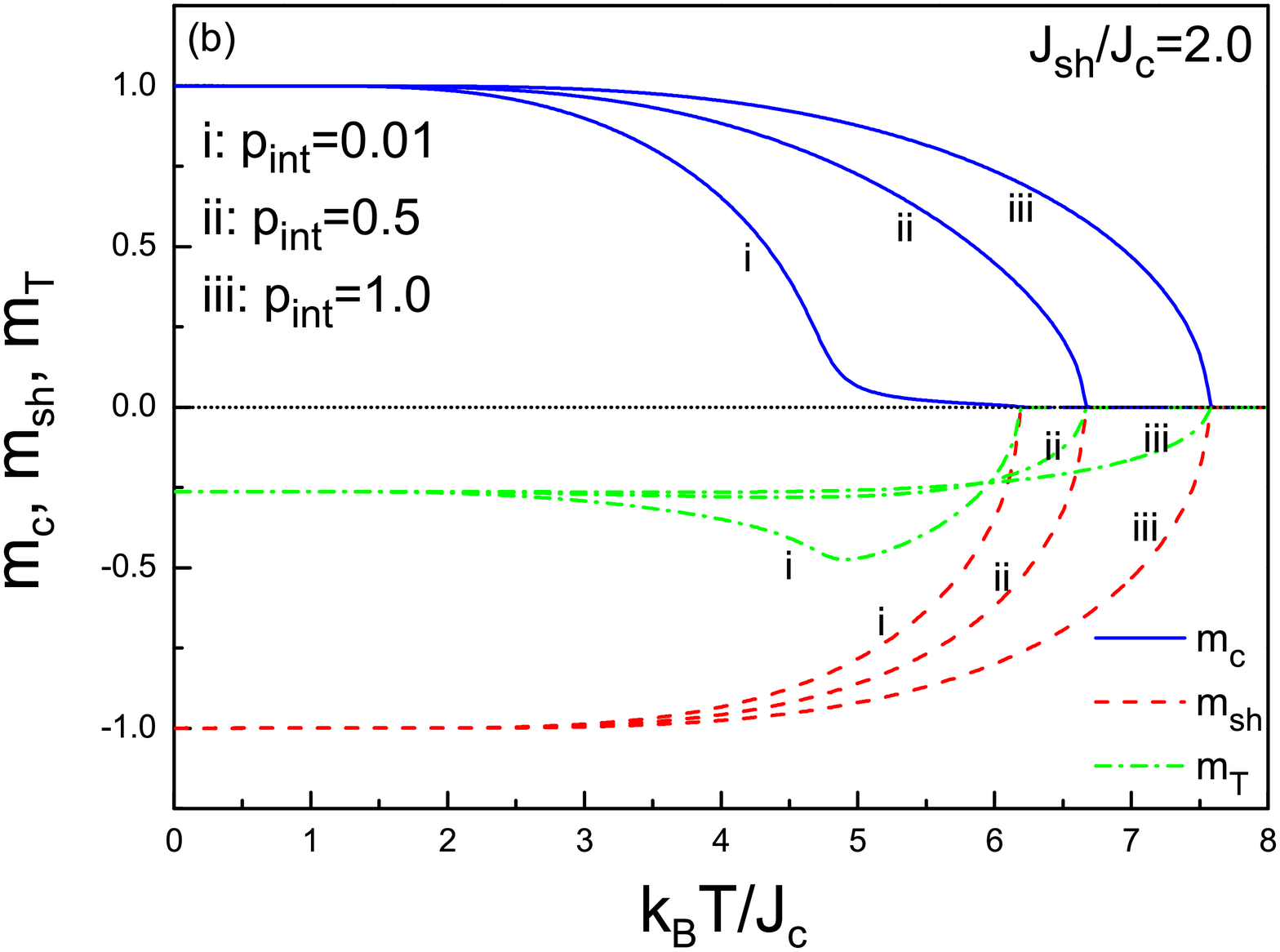}\\
\caption{Magnetization curves corresponding to the phase diagrams in Fig. \ref{fig8} with some selected values of interface bond concentration $p_{int}$ for (a) $J_{sh}/J_{c}=0.5$, and (b) $J_{sh}/J_{c}=2.0$. The other parameters are fixed as $J_{int}/J_{c}=-1.0$, $p_{c}=1.0$ and  $p_{sh}=1.0$. Solid, dashed and dashed-dotted lines correspond to the core, shell, and total magnetization curves, respectively.}\label{fig9}
\end{figure}

In Fig. \ref{fig8}, dependence of the critical and compensation temperatures of the nanowire system on the ferromagnetic shell coupling constant $J_{sh}$ and interface bond concentration $p_{int}$ with $J_{int}/J_{c}=-1.0$ is depicted for selected two values of the interface bond concentration. As seen in Fig. \ref{fig8}a, when the interface bonds are completely diluted, compensation point depresses to zero at $J_{sh}/J_{c}=0.0$ whereas for completely pure interface bonds, it has a finite value even at $J_{sh}/J_{c}=0.0$. In both cases, as $J_{sh}/J_{c}$ increases then compensation point of the system increases, and after a specific value of $J_{sh}/J_{c}$, it disappears and critical temperature of the system increases almost linearly. We note that for strongly disordered interface bonds (i.e. $p_{int}=0.0$) the system exhibits similar characteristics of a semi-infinite ferromagnet \cite{kaneyoshi_final}. Critical temperature curves in Figs. \ref{fig8}a and \ref{fig8}b are identical to those obtained in Fig. \ref{fig2}a for a ferromagnetic nanowire system. Magnetization curves corresponding to the phase diagrams shown in Fig. \ref{fig8} are depicted in Fig. \ref{fig9}. In Fig. \ref{fig9}a, we can clearly see that when the shell coupling constant $J_{sh}$ is relatively weaker than the core coupling $J_{c}$ then the transition temperature of the system is governed by the core layer of the nanowire at low $p_{int}$ values. Contribution of the shell layer becomes significant with increasing $p_{int}$ values. On the other hand, when the shell coupling is greater than that of the core, the mechanism works in an opposite way. Namely, as shown in Fig. \ref{fig9}b where $J_{sh}/J_{c}=2.0$, at low $p_{int}$ values, transition temperature of the system is determined by the shell layer,  while core layer contributes to the transition temperature as $p_{int}$ increases. Another interesting point in Fig. \ref{fig9} is the classification of the total magnetization curves of the antiferromagnetic nanowire system. In the bulk ferrimagnetism of N\'{e}el \cite{neel,strecka}, it is possible to classify the thermal variation of the total magnetization curves ($m_{T}$) in five categories. The curves in Fig. \ref{fig9}a exhibit N-type behavior whereas in Fig. \ref{fig9}b, we see P-type and Q-type dependencies with negative saturation magnetizations.

\begin{figure}[!h]
\center
\includegraphics[width=8cm]{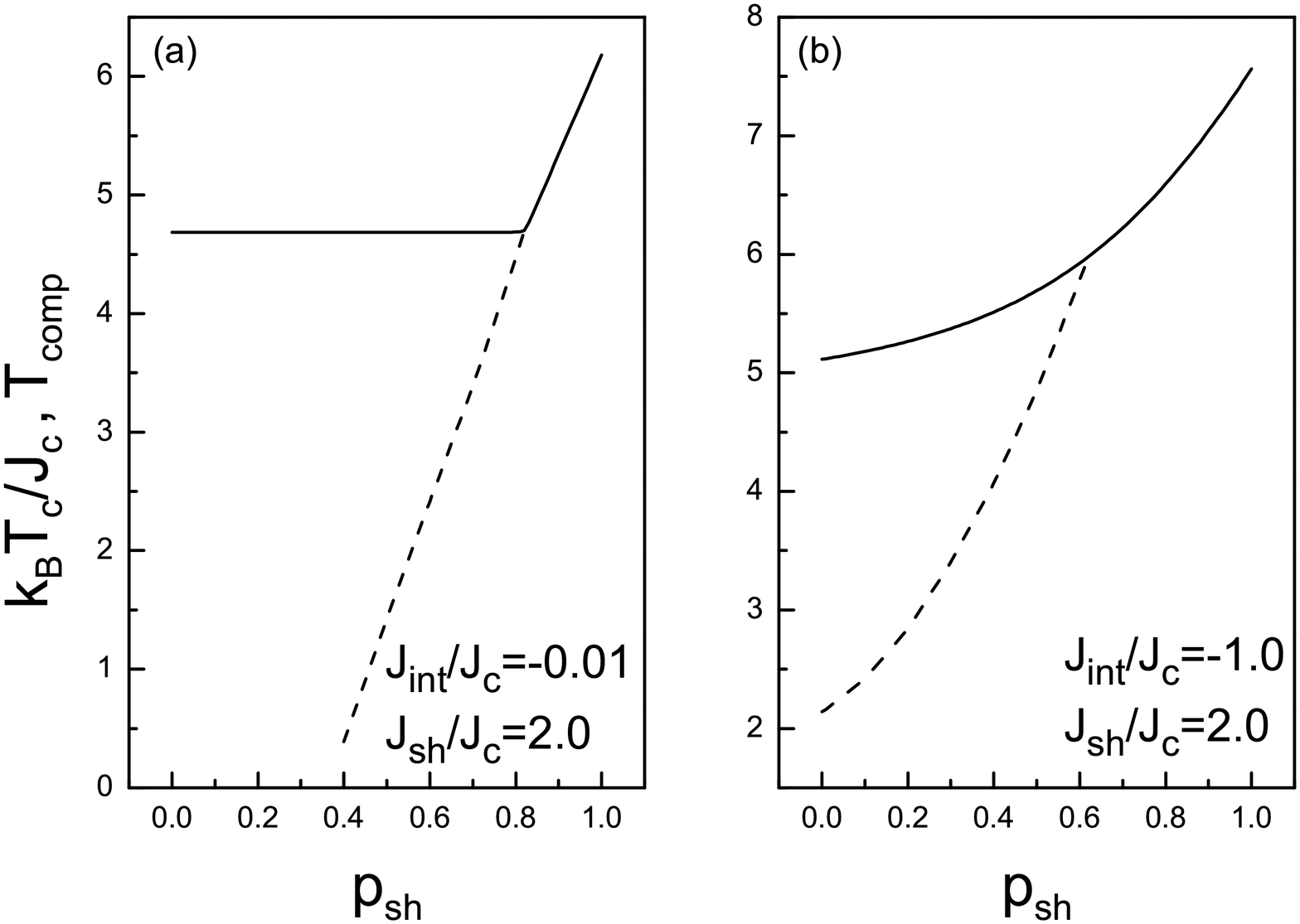}\\
\caption{Variation of critical and compensation temperatures of the antiferromagnetic nanowire system as a function of concentration of ferromagnetic shell bonds $p_{sh}$ for (a) $J_{int}/J_{c}=-0.01$, (b) $J_{int}/J_{c}=-1.0$. In both panels, we set $J_{sh}/J_{c}=2.0$, $p_{int}=1.0$ and $p_{c}=1.0$. Solid (dashed) line represent the transition temperature (compensation point) of the system.}\label{fig10}
\end{figure}
\begin{figure}[!h]
\center
\includegraphics[width=8cm]{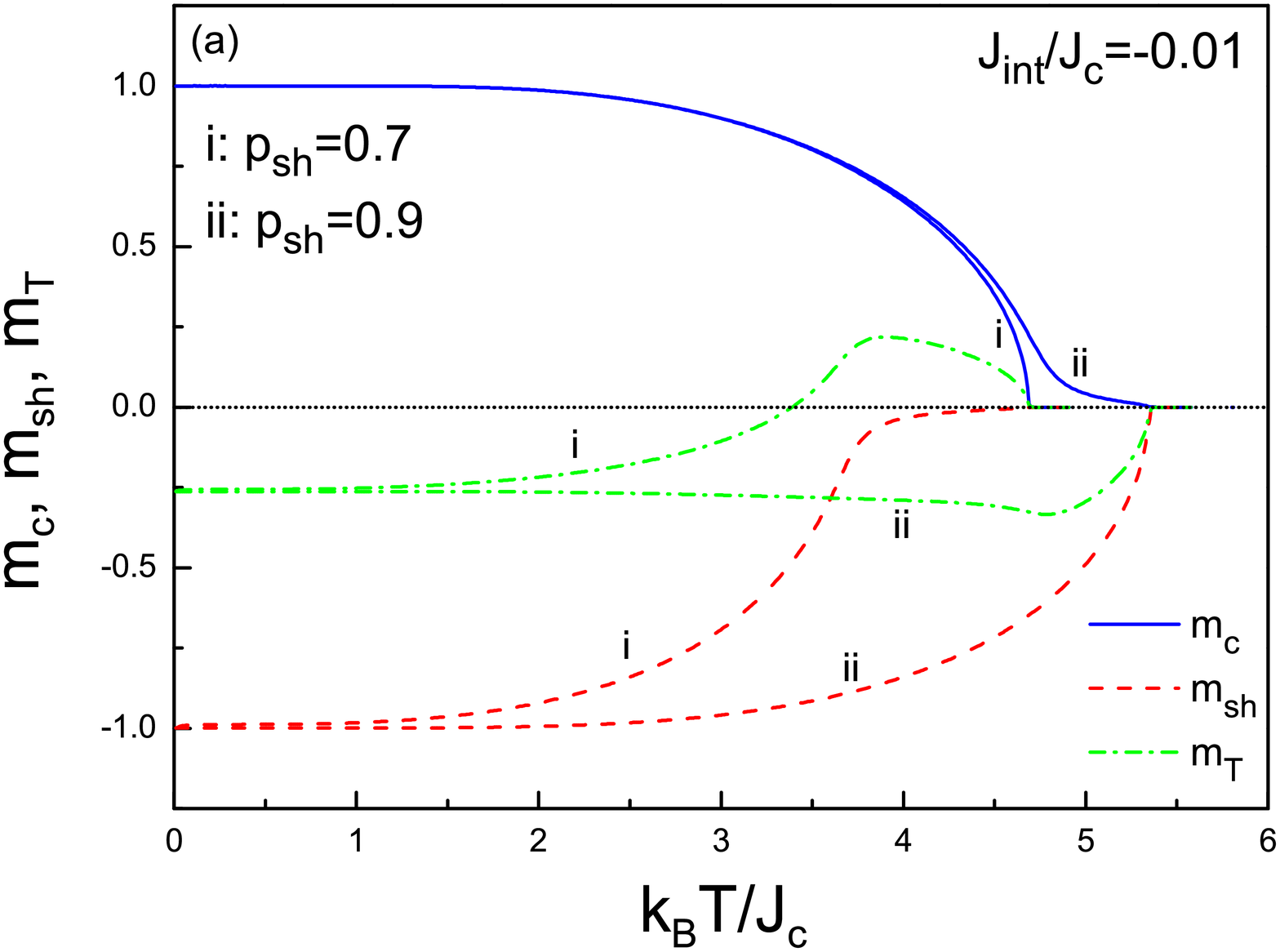}
\includegraphics[width=8cm]{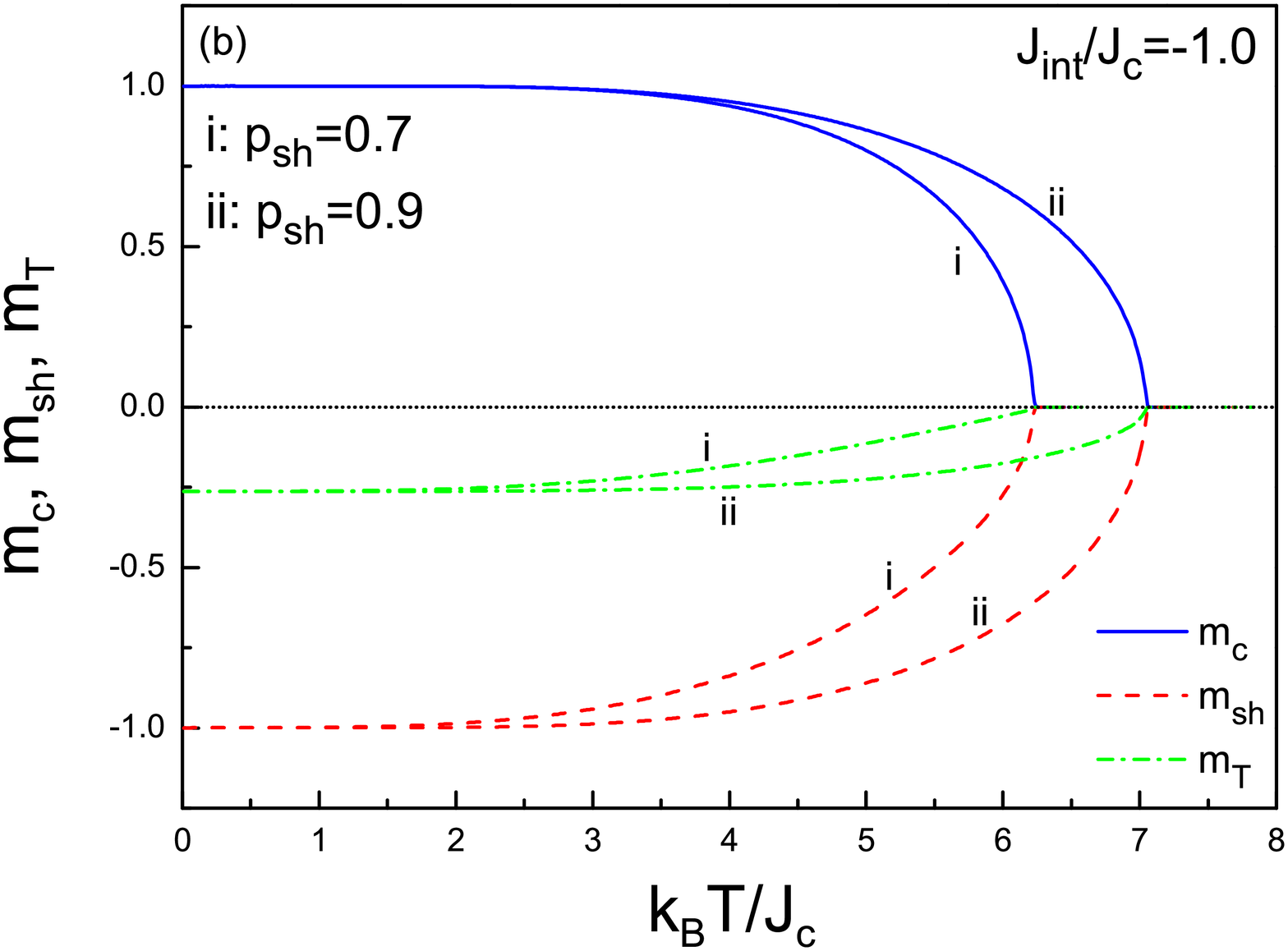}\\
\caption{Magnetization curves corresponding to the phase diagrams in Fig. \ref{fig10} with some selected values of bond concentration in the shell layer $p_{sh}$ for (a) $J_{int}/J_{c}=-0.01$, and (b) $J_{int}/J_{c}=-1.0$. The other parameters are fixed as $J_{sh}/J_{c}=2.0$, $p_{int}=1.0$, $p_{c}=1.0$. Solid, dashed and dashed-dotted lines correspond to the core, shell, and total magnetization curves, respectively.}\label{fig11}
\end{figure}

In order to investigate the effect of the concentration of disordered bonds in the shell layer on the antiferromagnetic properties of the system, we represent the variation of critical and compensation temperatures with $p_{sh}$ in Fig. \ref{fig10} for $p_{int}=1.0$ and $p_{c}=1.0$. In Figs. \ref{fig10}a and \ref{fig10}b, the situation is depicted for weak and strong core-shell interface interactions, respectively. Critical temperature curves in Figs. \ref{fig10}a and \ref{fig10}b are identical to the curves labeled (i) in Figs. \ref{fig2}d and \ref{fig2}a, respectively corresponding to the ferromagnetic system. Behavior of the critical temperature curves in Figs. \ref{fig10}a and \ref{fig10}b are already explained before, therefore we will not discuss them here. In Figs. \ref{fig10}a and \ref{fig10}b, we see some novel phenomena, namely the appearance of a compensation point induced by the surface dilution. That is, if the exchange interactions in the shell layer are completely pure then we observe only a critical temperature in the system. However, as the concentration of the active bonds in the shell layer is decreased, we can understand that the system may
exhibit a compensation point which depresses to zero at a critical concentration $p_{sh}^{\mathrm{crit.}}$ for weak antiferromagnetic core-shell interactions, as depicted in Fig. \ref{fig10}a. As the strength of antiferromagnetic interactions between the core and shell spins increases, compensation point remains at a finite value, even for $p_{sh}=0.0$. Some selected magnetization curves corresponding to the phase diagrams in Fig. \ref{fig10} are depicted in Fig. \ref{fig11}. As seen in this figure, the magnetization curves exhibit N-type, P-type and Q-type behaviors, as shown in Figs. \ref{fig11}a and \ref{fig11}b,  respectively. P-type and Q-type curves exhibit negative saturation magnetization.

\begin{figure}[!h]
\center
\includegraphics[width=8cm]{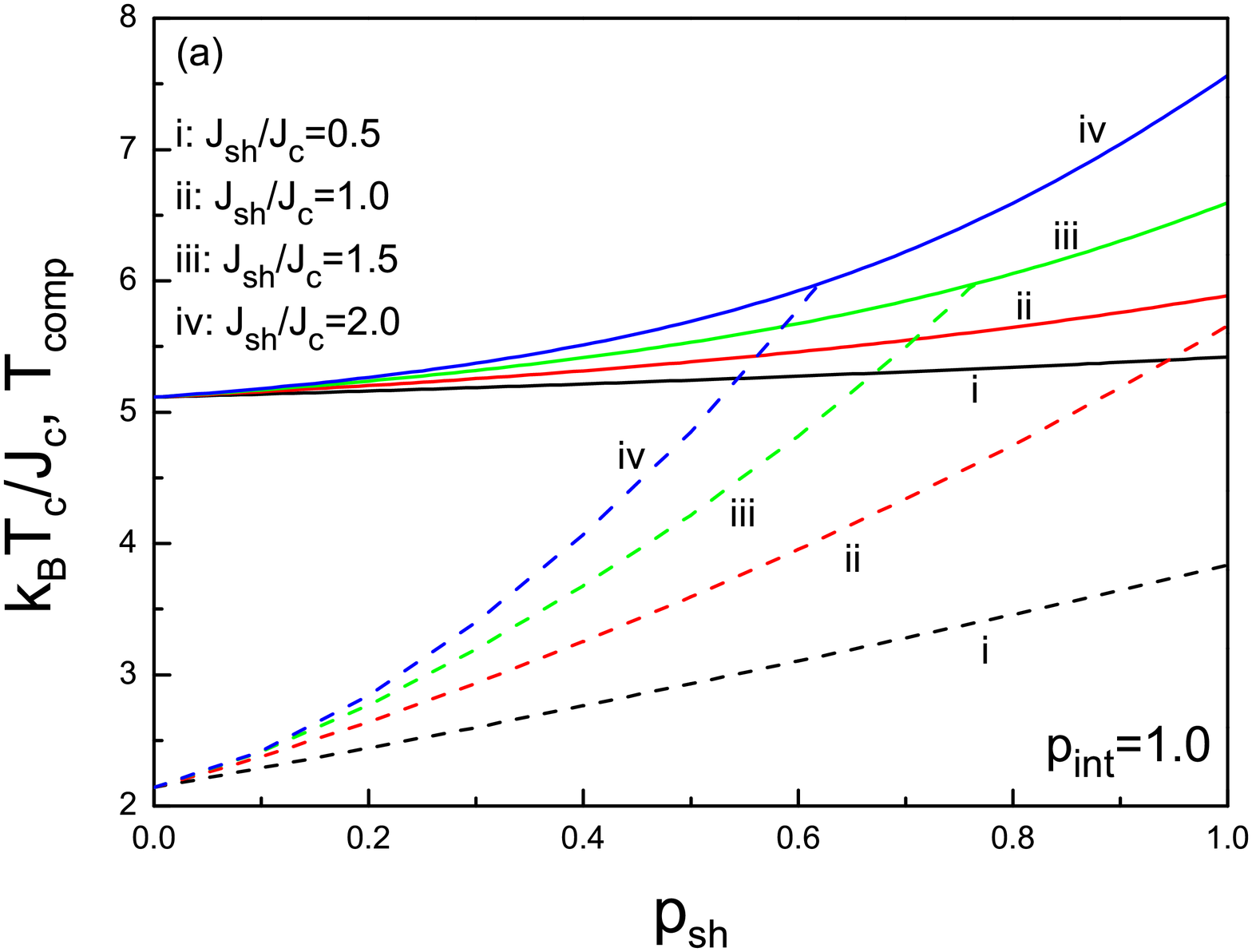}
\includegraphics[width=8cm]{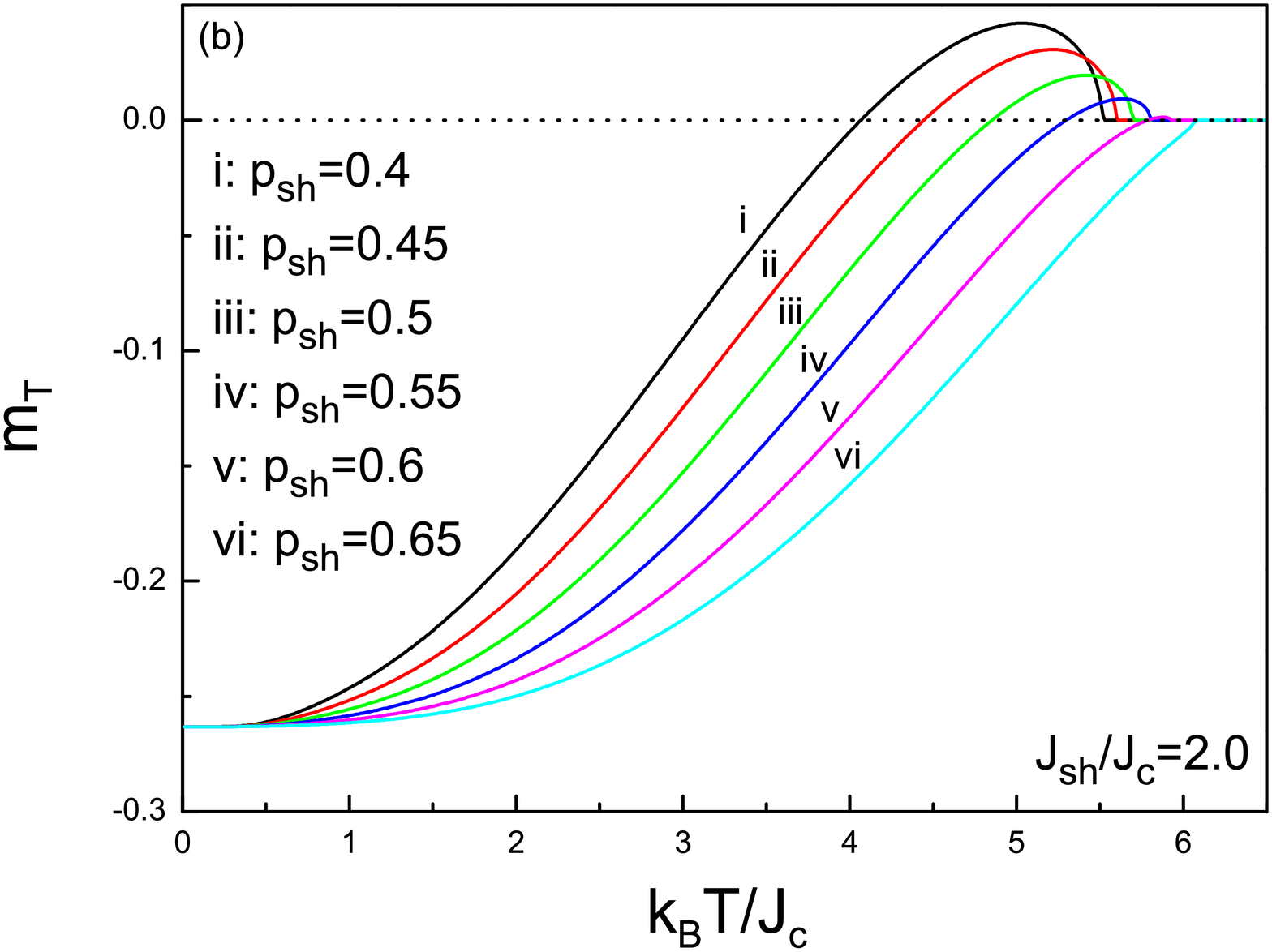}\\
\caption{(a) Variation of transition temperature and compensation point curves with dilution parameter $p_{sh}$ in the nanowire system with antiferromagnetic interface interactions. Solid (dashed) lines represent transition temperature (compensation point) curves. (b) Total magnetization curves for some selected values of $p_{sh}$ for $J_{sh}/J_{c}=2.0$. The other parameters are fixed as $J_{int}/J_{c}=-1.0$, $p_{int}=1.0$, $p_{c}=1.0$.}\label{fig12}
\end{figure}
\begin{figure}[!h]
\center
\includegraphics[width=8cm]{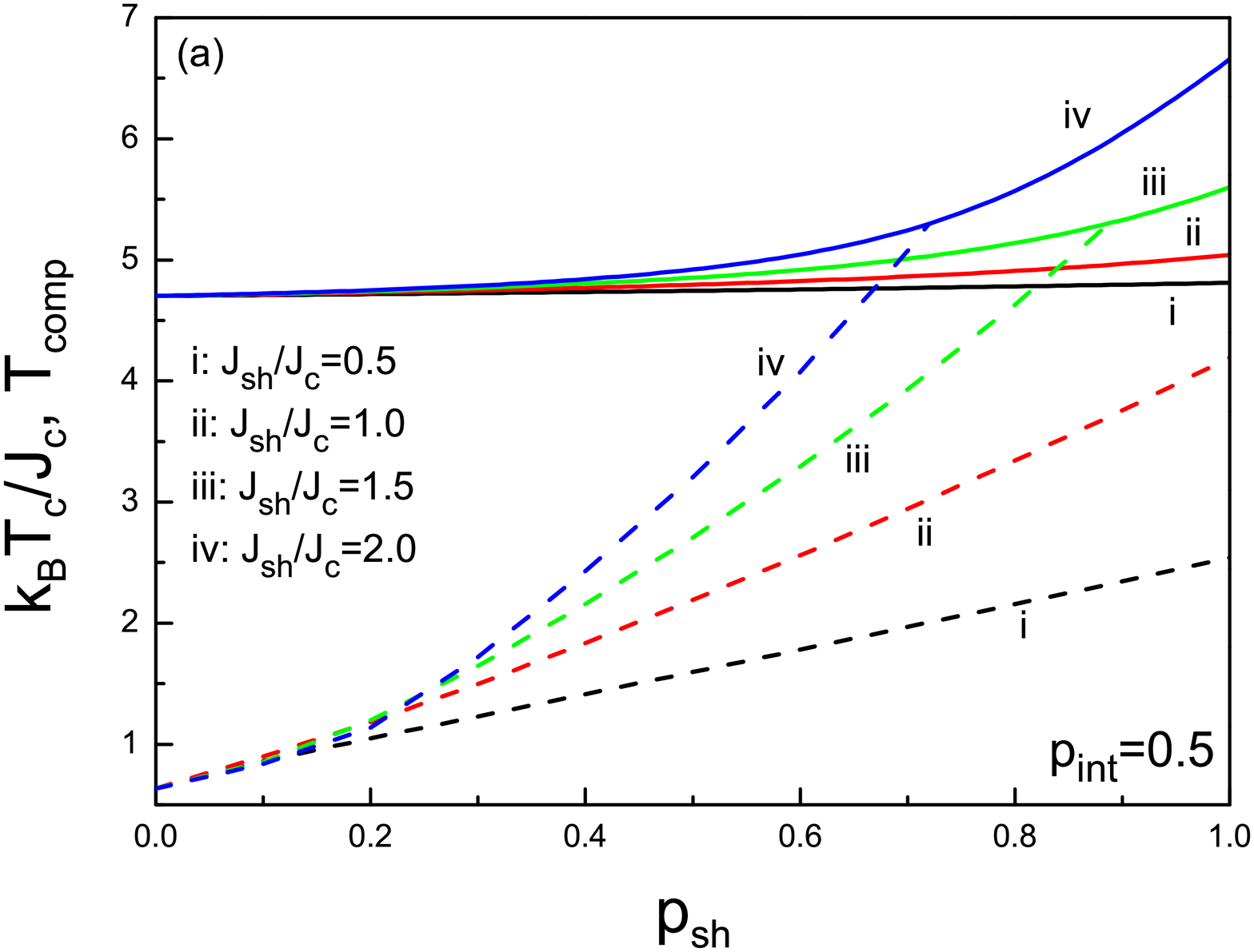}
\includegraphics[width=8cm]{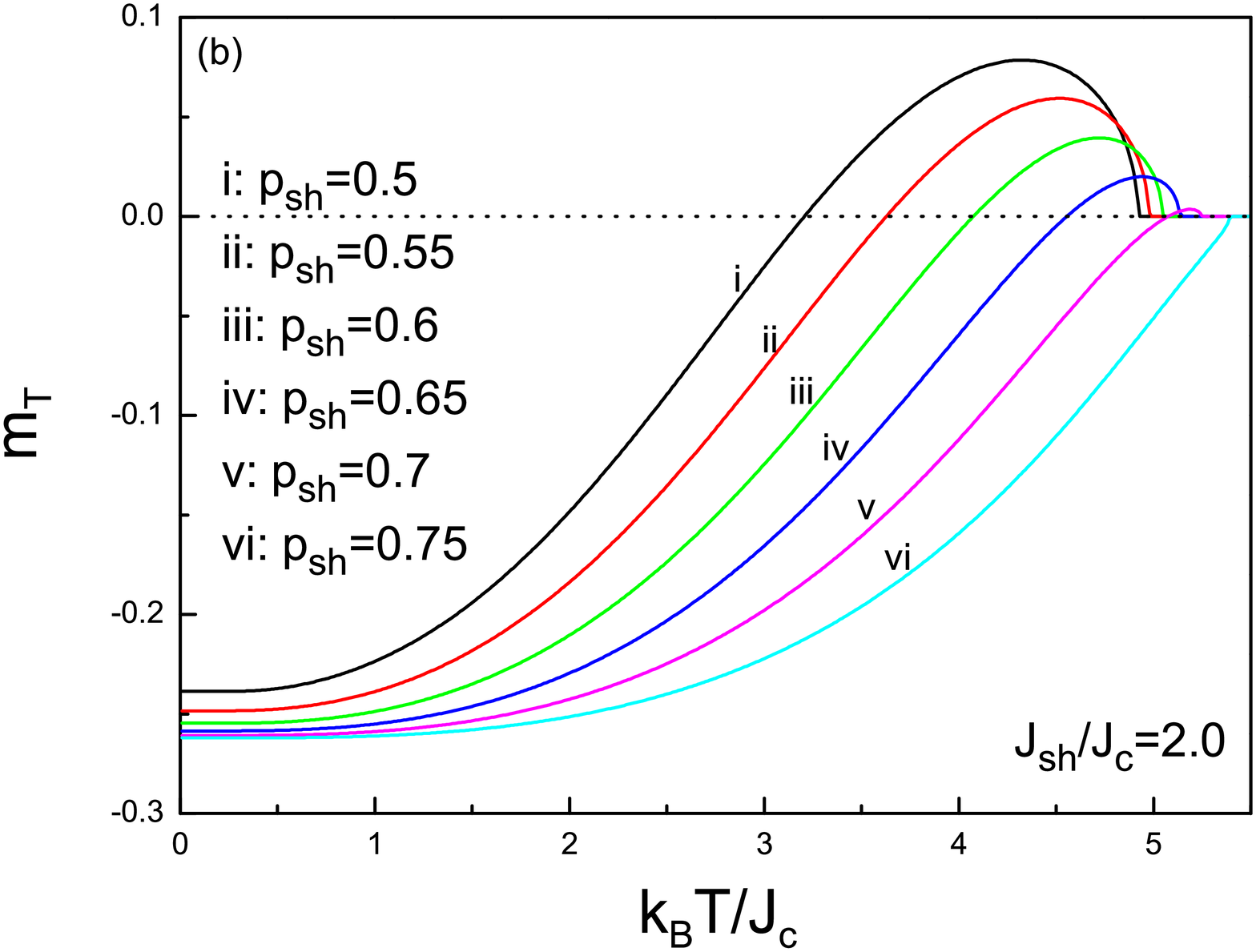}\\
\caption{(a) Variation of transition temperature and compensation point curves with dilution parameter $p_{sh}$ in the nanowire system with antiferromagnetic interface interactions. Solid (dashed) lines represent transition temperature (compensation point) curves. (b) Corresponding total magnetization curves for some selected values of $p_{sh}$ for $J_{sh}/J_{c}=2.0$. The other parameters are fixed as $J_{int}/J_{c}=-1.0$, $p_{int}=0.5$, $p_{c}=1.0$.}\label{fig13}
\end{figure}
\begin{figure}[!h]
\center
\includegraphics[width=8cm]{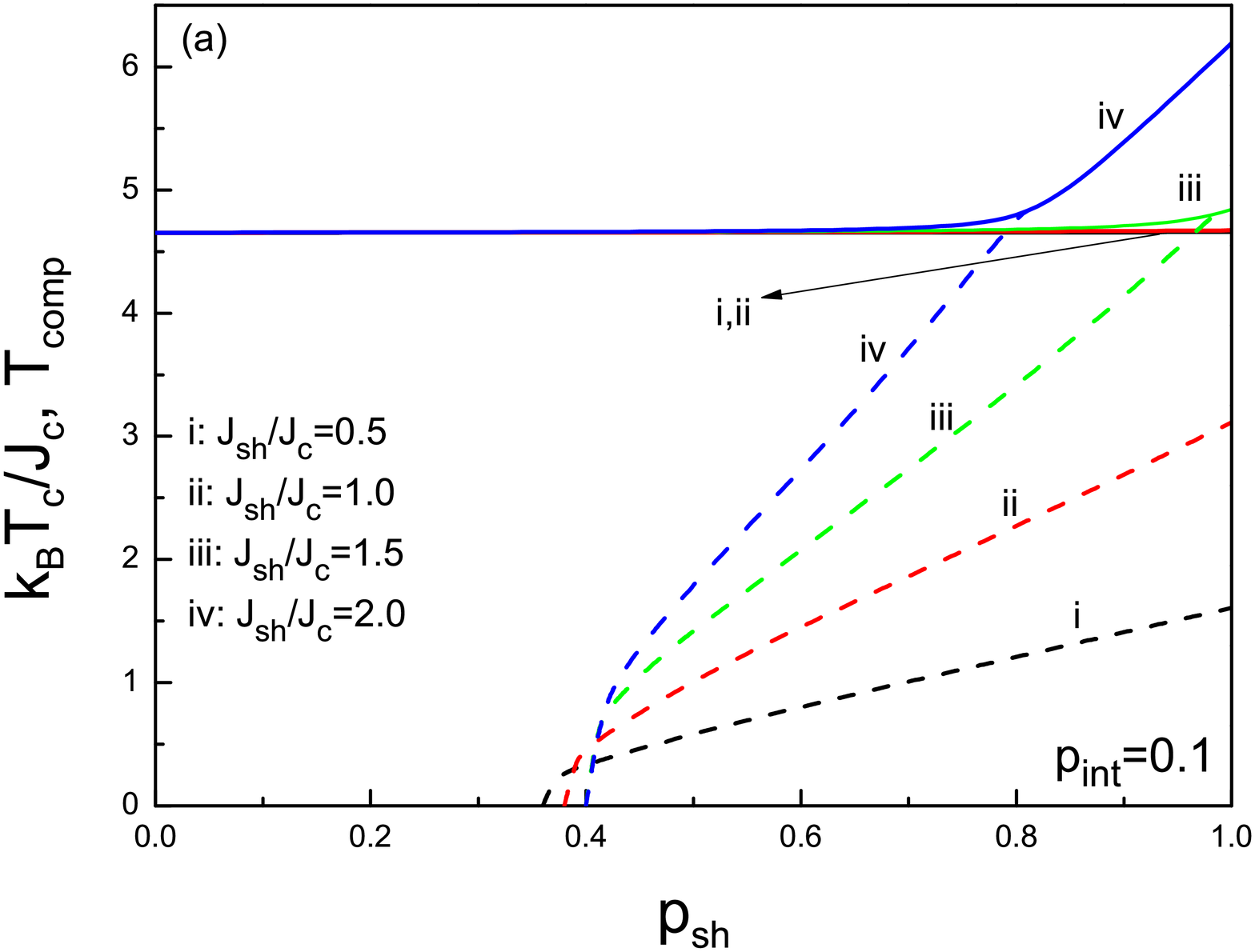}
\includegraphics[width=8cm]{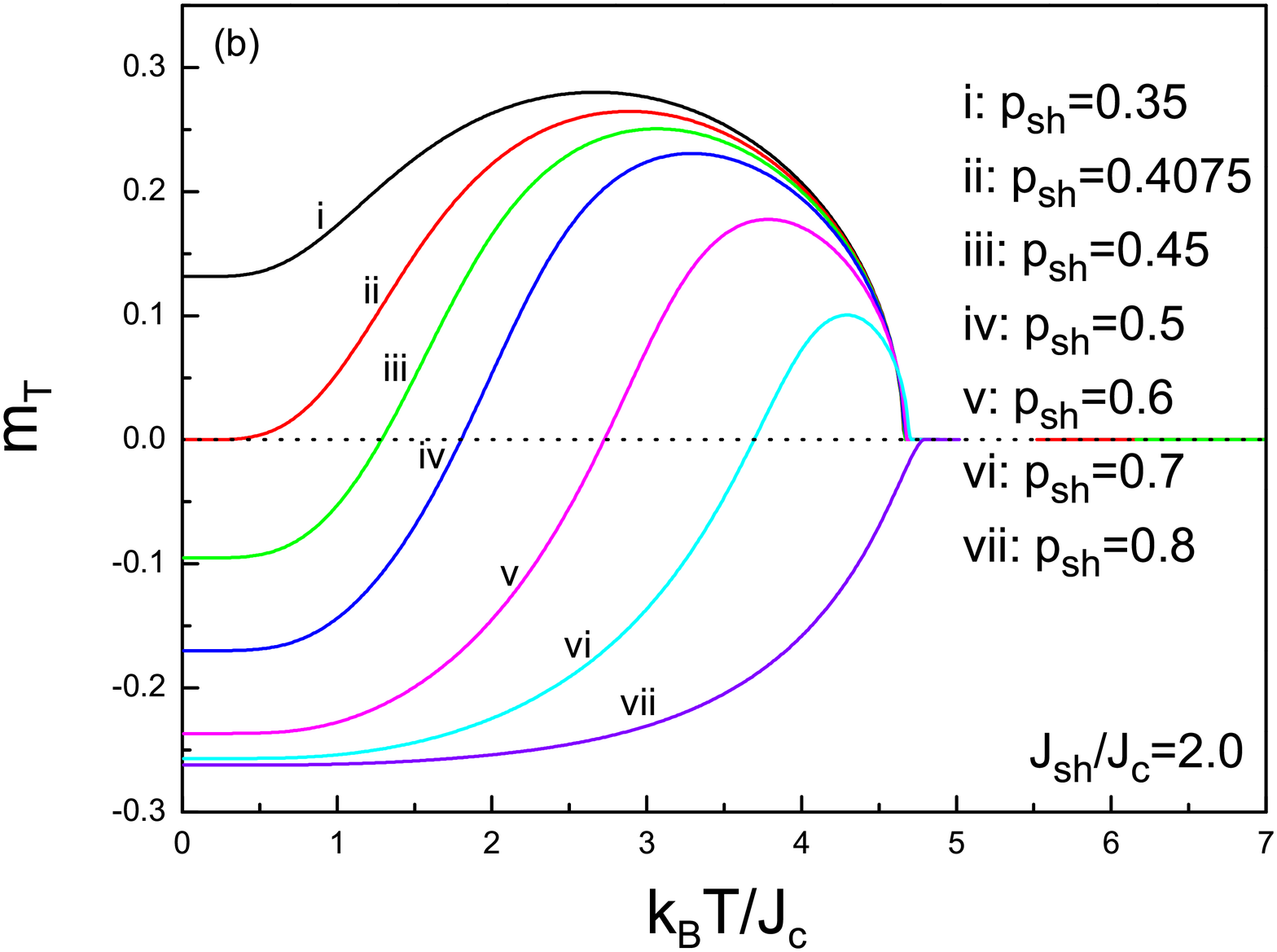}\\
\caption{(a) Variation of transition temperature and compensation point curves with dilution parameter $p_{sh}$ in the nanowire system with antiferromagnetic interface interactions. Solid (dashed) lines represent transition temperature (compensation point) curves. (b) Corresponding total magnetization curves for some selected values of $p_{sh}$ for $J_{sh}/J_{c}=2.0$. The other parameters are fixed as $J_{int}/J_{c}=-1.0$, $p_{int}=0.1$, $p_{c}=1.0$}\label{fig14}
\end{figure}

In a previously published work \cite{kaneyoshi4}, in order to clarify the influence of the site dilution in the surface shell on the transition temperature and compensation point of the antiferromagnetic nanowire system, the phase diagrams have been investigated for several values of the antiferromagnetic interface interaction strength $J_{int}/J_{c}$, and it has been found that all of the compensation point curves reduce to zero at a critical site concentration which has a value $p_{\mathrm{site}}^{*}=0.583$ (see Fig. \ref{fig2} in Ref. \cite{kaneyoshi4}). In order to compare these results with the present work, we plot the phase diagrams in $(k_{B}T_{c}/J_{c}-p_{sh})$ plane for several values of $p_{int}$ and the corresponding magnetization curves in Figs. \ref{fig12}-\ref{fig14}. In Fig. \ref{fig12}a, we select $J_{int}/J_{c}=-1.0$, $p_{int}=1.0$ and $p_{c}=1.0$ which means only the ferromagnetic shell bonds are diluted. In this figure, it is clear that the compensation point always exists in the system even for $p_{sh}=0.0$. According to the magnetization curves shown in Fig. \ref{fig12}b, the system exhibits N-type and Q-type characteristics. When the interface bonds are somewhat diluted, such as the case $p_{sh}=0.5$ in Fig. \ref{fig13}a, again we observe a finite compensation point in the system for $p_{sh}=0.0$. Furthermore, the magnetization curves in Fig. \ref{fig13}b exhibit similar characteristics as in Fig. \ref{fig12}b. The only difference is that the saturation magnetizations in Fig.\ref{fig12}b change as $p_{sh}$ values are varied whereas in Fig. \ref{fig12}a saturation magnetizations are not altered by varying $p_{sh}$ values. According to our calculations, the reason is due to the fact that as the interface bonds are diluted, core and shell layers become independent of each other, and saturation magnetization of the shell layer begins to be affected by varying $p_{sh}$ values. Moreover, if there exists a strong disorder in the antiferromagnetic interface bonds between core and shell layers then all of the compensation point curves in Fig. \ref{fig14}a  reduce to zero at a critical concentration $p_{sh}^{\mathrm{crit.}}$. In this case, we found that the magnetization curves exhibit very rich characteristics. Namely, as seen in Fig. \ref{fig14}b, the magnetization curves exhibit P-type, L-type, N-type, and Q-type characteristics within the given range of $p_{sh}$ in Fig. \ref{fig14}b. The effect of the dilution parameter of the antiferromagnetic interface bonds $p_{int}$ can be clarified by comparing the phase diagrams in Figs. \ref{fig12}a-\ref{fig12}c. As is seen in Figs. \ref{fig12}a-\ref{fig12}c, the system exhibits a finite compensation point for $p_{int}=1.0$ and $p_{int}=0.5$ even for $p_{sh}=0.0$ whereas for $p_{int}=0.1$ there exist a critical concentration of shell bonds $p_{sh}^{\mathrm{crit.}}$ value of which depends on $J_{sh}/J_{c}$.

\section{Conclusion}\label{conclusion}
In conclusion, in order to clarify how the magnetism in a nanoparticle system is affected in the presence of disordered bonds in the surface shell and also in the interface between the core and shell layers of the particle, we have considered cylindrical nanowire systems with both ferromagnetic and antiferromagnetic interactions at the core-shell interface within the framework of EFT method. The system has been modeled by a pure ferromagnetic core which is surrounded by a ferromagnetic shell layer. In order to consider the effects of the quenched disordered bonds (i.e. bond dilution), particle shell, and also interface exchange couplings are randomly  distributed on the lattice, according to given certain distribution functions.

A complete picture of the phase diagrams and magnetization profiles have been represented. It has been shown that for the antiferromagnetic nanowire system, the magnetization curves can be classified according to N\'{e}el theory of ferrimagnetism and it has been found that under certain conditions, the magnetization profiles may exhibit Q-type, P-type, N-type and L-type behaviors. The observed L-type behavior has not been reported in the literature before for the equilibrium properties of nanoscaled magnets. Furthermore, as another interesting feature of the system, it has been found that a compensation point can be induced by a bond dilution process in the surface. Finally, we have not found any evidence of neither the first order phase transition characteristics, nor the reentrance phenomena.

\section*{Acknowledgements}
One of the authors (YY) would like to thank the Scientific and
Technological Research Council of Turkey (T\"{U}B\.{I}TAK) for
partial financial support. The numerical calculations reported in this paper were performed at T\"{U}B\.{I}TAK ULAKBIM, High Performance and Grid Computing Center (TR-Grid e-Infrastructure).


\end{document}